\documentclass[a4paper,fleqn,usenatbib]{mnras}
\usepackage{newtxtext,newtxmath}
\usepackage{changepage}
\usepackage[flushleft]{threeparttable}
\usepackage{rotating,booktabs}
\usepackage[T1]{fontenc}
\usepackage{ae,aecompl}
\usepackage{epsfig}
\usepackage{amsmath, amsfonts, epsfig, xspace}
\usepackage{natbib}
\usepackage{rotating}
\usepackage{graphicx}
\usepackage{caption}
\usepackage{longtable}
\usepackage{multicol}
\usepackage{booktabs}
\usepackage{amsmath}
\usepackage{rotating}
\usepackage{booktabs}
\usepackage{changepage}
\usepackage{subfig}
\usepackage{hyperref}
\usepackage{multirow}

\usepackage{textcomp}
\usepackage{scalerel}
\def\kms{~kms$^{-1}$}
\newcommand{\sqcm}{cm$^{-2}$}

\newcommand{\Msun}{$\rm M_{\odot}$}

\newcommand{\rhocl}{$\rho_{\rm cl}$} 
\newcommand{\zcl}{$z_{\rm cl}$} 
\newcommand{\nrhocl}{$\rho_{\rm cl}/R_{\rm 500}$}

\def\civab{C~{\sc iv}~$\lambda\lambda$1548,1550~}

\def\lya{H\,{\sc i}~$\lambda$1215} 
\def\ovi{O\,{\sc vi}~$\lambda$1031} 
 
\def\oviab{O~{\sc vi}$\lambda\lambda$1031,1037} 

\newcommand{\HI}{\mbox{H\,{\sc i}}}

\newcommand{\MgII}{\mbox{Mg\,{\sc ii}}}
\newcommand{\FeII}{\mbox{Fe\,{\sc ii}}}
\newcommand{\SiII}{\mbox{Si\,{\sc ii}}}
\newcommand{\SiIII}{\mbox{Si\,{\sc iii}}}
\newcommand{\SiIV}{\mbox{Si\,{\sc iv}}}
\newcommand{\CII}{\mbox{C\,{\sc ii}}}
\newcommand{\CIII}{\mbox{C\,{\sc iii}}}
\newcommand{\CIV}{\mbox{C\,{\sc iv}}}
\newcommand{\OVI}{\mbox{O\,{\sc vi}}}

\newcommand{\Lya}{Ly$\alpha$}
\newcommand{\Lyb}{Ly$\beta$}

\def\siglya{$\sigma_{\scaleto{\rm v,~\HI \rm}{5pt}}$}

\newcommand{\ewlya}{$W_{r}^{1215}$}

\newcommand{\ewovi}{$W_{r}^{1031}$}
\newcommand{\ewciv}{$W_{r}^{1548}$}
\def\ewmgiia{W$_{r}^{2796}$}
\def\ewmgiia{W$_{r}$(2796)}
\def\aj{{AJ}}%
\def\araa{{ARA\&A}}%
\def\apj{{ApJ}}%
\def\apjl{{ApJ}}%
\def\apjs{{ApJS}}%
\def\aap{{A\&A}}%
\def\mnras{{MNRAS}}%
\title[Gas and metals in cluster outskirts]{Characterizing cool, neutral gas and  ionized metals in the outskirts of low-$z$ galaxy clusters} 
\author[Mishra et al.]{{\Large Sapna Mishra$^{1}$, Sowgat Muzahid$^{1}$, Sayak Dutta$^{1}$, Raghunathan  Srianand$^{1}$, and Jane Charlton$^{2}$} \\\\
$^{1}$IUCAA, Post Bag 04, Ganeshkhind, Pune -- 411007, India \\  
$^{2}$Department of Astronomy \& Astrophysics, The Pennsylvania State University, 525 Davey Lab, University Park, PA 16802, USA 
}

\begin{document}
\date{Accepted ---. Received ---; in original form ---}

\pagerange{\pageref{firstpage}--\pageref{lastpage}} \pubyear{2021}

\maketitle

\label{firstpage}
\begin{abstract}

\noindent 
We present the first detection of cool, neutral gas in the outskirts of low-$z$ galaxy clusters using a statistically significant sample of 3191 $z\approx 0.2$ background quasar--foreground cluster pairs by cross-matching the Hubble Spectroscopic Legacy Archive quasar catalog with optically- and SZ-selected cluster catalogs. The median cluster mass of our sample is $\approx 10^{14.2}$~\Msun\, with a median impact parameter (\rhocl) of $\approx5$~Mpc. 
We detect significant \Lya, marginal \CIV, but no \OVI\ absorption in the signal-to-noise ratio weighted mean stacked spectra with rest-frame equivalent widths of $0.096\pm0.011$~\AA, $0.032\pm0.015$~\AA, and $<0.009$~\AA\ ($3\sigma$) for our sample.   
The \Lya\ REW shows a declining trend with increasing \rhocl\ (\nrhocl) which is well explained by a power-law with a slope of $-0.79$ ($-0.70$). The covering fractions (CFs) measured for \Lya\ (21\%), \CIV\ (10\%) and \OVI\ (10\%) in cluster outskirts are significantly lower than in the circumgalatic medium (CGM). 
We also find that the CGM of galaxies that are closer to cluster centers or that are in massive clusters is considerably deficient in neutral gas.
The low CF of the \Lya\ along with the non-detection of \Lya\ signal when the strong absorbers ($N(\HI) > 10^{13}$~\sqcm) are excluded, indicate the patchy distribution of cool gas in the outskirts. We argue that the cool gas in cluster outskirts in combination arises from the  circumgalactic gas stripped from cluster galaxies and to large-scale filaments feeding the clusters with cool gas.
\end{abstract}

\section{Introduction}
\label{sec:intro}

According to the standard structure formation model, galaxy clusters, being the most massive structure in the universe, mark the nodal points in the cosmic web where several filamentary strands intersect. The inflow of cosmic matter via these filaments feeds the growth of galaxy clusters. The infalling gas in the  intracluster medium (ICM i.e., $<R_{500}$\footnote{Radius within which the mean mass density of a cluster is 500 times the critical density of the universe. Similarly,  $R_{200}$ corresponds to the radius at which the mean mass density of a cluster is 200 times the critical density of the universe.}), having been shock-heated to a very high temperature of $T\sim$ 10$^{7-8}$~K \citep{Dave1999,Voit2005}, is bright enough to be directly mapped in X-ray observations \citep{Urban2014,Simionescu2015,Bifi2018}. This hot phase of gas in the ICM is thought to account for 80\% of the baryonic content in the clusters. On the other hand, clusters are constantly growing and evolving in their outskirts ($>R_{500}$) as a result of a succession of galaxy mergers and the accretion of infalling gas from the intergalactic medium (IGM). Interestingly, despite the fact that a significant portion of the IGM is in the cool/warm phase \citep[$T\sim$ 10$^{4-5}$~K;][]{Dave2010,Kravtsov2012}, the nature of this gas phase remains poorly understood in the outskirts of clusters. Given the lack of sensitive X-ray diagnostics for directly probing this gas in emission, absorption line spectroscopy of UV-bright background quasars can be leveraged as an ideal alternative to study this otherwise invisible yet crucial phase of the cluster outskirts. \par

There are a handful of studies focusing on the distribution of neutral gas traced by \Lya\ but using limited numbers of quasar sightlines,  primarily using Hubble Space Telescope Cosmic Origin Spectrograph ($HST$/COS) spectra of background quasars \citep[e.g.,][]{yoon2012,Tejos2016,Muzahid2017,Yoon2017,Burchett2018}, but see \citet[]{Lanzetta1996,Tripp1998,Miller2002} for pre-COS studies. Using 23 quasar sightlines in the background of the Virgo cluster, \citet{yoon2012} found that \Lya\ absorbers avoid the hot ICM and are more abundant in the outskirts. Based on the  concomitant occurrence of these absorbers with \HI-emitting substructures in their study, the authors posited that the warm gas is tracing the large-scale structure (LSS). In a subsequent study of 29 and 8 quasar sightlines passing through the Virgo and Coma clusters respectively, \citet{Yoon2017} concluded that \Lya\ absorbers are more prevalent between 1$-$2 $R_{\rm vir}$\footnote{where R$_{\rm vir}$ in their study is defined at the overdensity of 100.} distance from the Virgo cluster center, while no such trend is observed for the Coma cluster. \citet{Burchett2018} also studied the outskirts (0.2$-$2.4 R$_{200}$) of 5 X-ray-selected clusters. They reported seven $N(\HI) > 10^{13}$~\sqcm\ \Lya\ absorbers in 4/5 clusters. Detection of significantly stronger \HI\ absorbers (i.e., $N(\HI) > 10^{16.5}$~\sqcm) is reported in the outskirts (1.6$-$4.7 $R_{500}$) of 3/3 SZ-selected clusters studied by \citet{Muzahid2017}.

Recently, \citet{Mishra2022}, using a large sample of $\approx80,000$ quasar$-$cluster pairs at $z\approx0.5$ from the Sloan Digital Sky Survey (SDSS), reported a detection of significant \MgII\ ($7\sigma$) and marginal \FeII\ ($3\sigma$) absorption in the mean and median stacked spectra of the quasars. From the density and metallicity constraints in their study, the authors suggested that the absorption signal is likely originating from the stripped gas from infalling galaxies. The authors further argued that the stripping process can be effective up to a distance of $\approx2.4$~Mpc ($\approx 3.6 R_{500}$) from the clusters. Additionally, \citet{anand2022cool} reported detection of \MgII\ absorption in the outskirts of clusters from the Dark Energy Spectroscopic Instrument (DESI) survey. They reported a covering fraction (CF) of 1--5\% within $R_{500}$ for \ewmgiia~$>0.4$~\AA, and concluded that the \MgII\ absorption likely stems from stripped interstellar medium (ISM) and/or satellite galaxies, based on the lack of correlations between the absorbers and properties of nearest cluster galaxies within $R_{200}$.\par

The nature of neutral gas in the cluster outskirts has also been explored in a handful of theoretical studies. For example, \citet{Emerick2015}, using hydrodynamical simulations, examined the distribution of \Lya\ absorbers around a Virgo-like and a Coma-like cluster. The authors found that the majority of their fast-moving low column density \Lya\ absorbers in the outskirts have filamentary origin. In the inner region, however, the increased column density and metallicity of the \Lya\ absorbers imply that gas from galaxies has been stripped away. To investigate the interplay between the ICM, cluster outskirts, and circumgalactic medium (CGM) of cluster galaxies, \citet{Butsky2019}, using a high-resolution hydrodynamical simulation, mapped the distribution of cool and warm gas around a $\sim$10$^{14}$~\Msun\ cluster. They found that, compared to the ICM, the cluster outskirts are more multiphased and richer in cool/warm gas. Due to the inadequate mixing of the ICM gas with the stripped gas from galaxies, the metallicity of the cool/warm gas phase in the outskirts shows a huge scatter compared to the metallicity of the hot ICM gas. In addition, they found the signature of stripping of the CGM of cluster galaxies out to $\approx 4 R_{200}$. \par

Cluster outskirts are ideal test-beds to study the environmental effects of cluster galaxies. It is well established both from observation and simulations that galaxies in dense environments, such as groups and clusters, are gas-deficient with elliptical morphology and redder colors than their field counterparts of comparable stellar mass \citep{Davies1973,Bosch2008,Wetzel2012,Bahe2013,Fossati2017,Davies2019,Hough2023,Kim2023,rohr2023}. However, the effects of cluster environments  on the CGM of galaxies, which is relatively loosely bound to galaxies as compared to the disc and the ISM, are not well explored observationally with statistically significant samples. \citet{Yoon2013} and \citet{Burchett2018} reported a lower CF of \Lya\ absorbers in the CGM of cluster galaxies compared to the field galaxies up to an impact parameter of 500~kpc from the galaxies. This trend has even been observed on group scales, with galaxies in groups showing an intermediate CF between cluster and field galaxies \citep[]{Burchett2018}. Interestingly, the simulation study by \citet{Bahe2013} has revealed that the ram pressure exerted by the extended halos of clusters can efficiently strip the hot gas from the infalling galaxies up to $\approx 5 R_{200}$. The authors found that other processes such as ``overshooting''  \citep{Gill2005} and ``pre-processing'' \citep{McGee2009} also play a role in stripping, depending on the mass of the infalling galaxy, its distance from the cluster core, and the cluster mass.

Here we present the first systematic study to probe and characterize the cool, neutral gas and ionized metals in the outskirts of low-$z$ clusters using a statistically significant sample of clusters. The purpose of this study is twofold: (i) to map the distribution of cool gas and metals surrounding clusters out to $10 R_{500}$, and (ii) to investigate the environmental effects on the CGM of cluster galaxies. We employ spectral stacking analysis to determine the average absorption strengths of the lines of interest. The major advantage of using spectral stacking is that this approach does not rely on any linking velocity to associate an absorber with a cluster, but at the expense of gas kinematics. However, in order to determine the covering fraction, we do use a linking velocity of $\pm500$~\kms\ (as generally used in the literature) for the visually identified absorbers.

The paper is structured as follows: In Section~\ref{sec:sample}, we discuss the quasar and cluster samples used in this study, followed by the construction details of the quasar-cluster pairs and the continuum normalisation of the HST/COS spectra. The key results are provided in Section~\ref{sec:results}. In Section~\ref{sec:discussion}, we discuss the nature and origin of \HI\ gas and metals detected in this work followed by the conclusion in Section~\ref{sec:conlusion}. We used a flat $\Lambda$CDM cosmology with $H_{0} =$ 71 km s$^{-1}$ Mpc$^{-1}$ , $\Omega_{\rm M} =$ 0.3, and $\Omega_{\Lambda} =$ 0.7.

\begin{figure*}
 \includegraphics[width=1.0\textwidth]{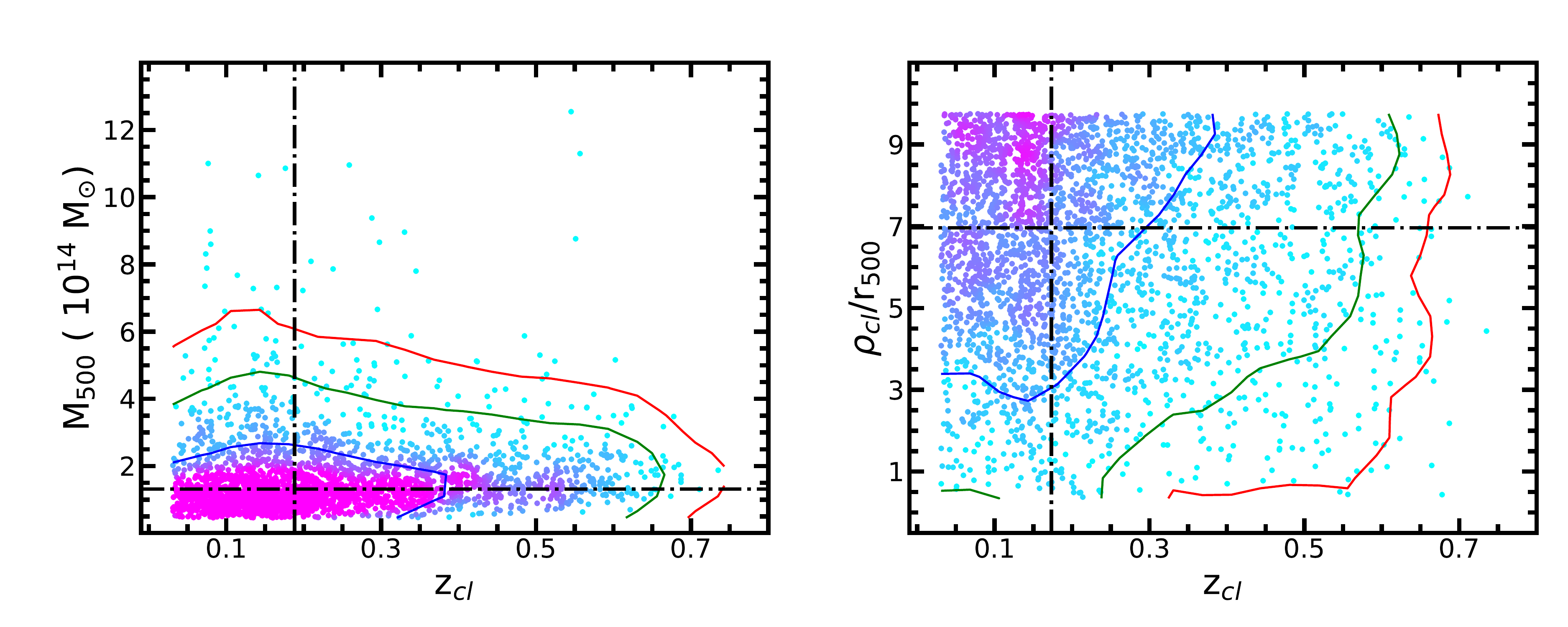}
 \caption{\emph{Left:} $M_{500}$ versus \zcl\ for the 2785 unique clusters. \emph{Right:} \nrhocl\ versus \zcl\ for the 3191 quasar-cluster pairs. The contours along with the density map in blue, green, and red respectively indicate the 68, 95, and 99.9 percentile enclosed regions. The black dashed-dotted horizontal and vertical lines show the median of the abscissa and ordinate parameters, respectively.}
 \label{fig:sample} 
\end{figure*}

\section{Sample}
\label{sec:sample}

\subsection{Cluster sample}  
\label{subsec:cluster-sample} 

To build a statistically significant sample of quasar$-$cluster pairs suitable for detecting the \Lya\ and UV metal absorption lines, we use seven cluster catalogs presented in (i) \citet[][hereafter BL15]{Bleem2015}, (ii) \citet[][hereafter WH15]{Wen2015}, (iii) \citet[][hereafter WHF18]{Wen2018}, (iv) \citet[][hereafter BL20]{Bleem2020}, (v) \citet[][hereafter H20]{Huang2020}, (vi) \citet[][hereafter H21]{Hilton2021}, and (vii) \citet[][hereafter Z21]{Zou2021}. As different algorithms are used to identify clusters in these catalogs, their identification criteria have varying limits on redshift, mass, and signal-to-noise ratio (SNR) for which their identification algorithms are complete. Consequently, we have only considered clusters from each catalog that meet their respective completeness criteria. The details on mass, redshift range covered, and completeness criteria of these catalogs are given in Appendix~\ref{appendix:catalog_summary}. In addition, we restrict our sample to clusters from these catalogs that have a known spectroscopic redshift for the brightest member galaxy.\par

Merging these seven cluster catalogs resulted in a total of 247,844 spectroscopically confirmed clusters. Similar to \citet{Mishra2022}, to eliminate repeated entries with somewhat different redshifts and sky positions within these clusters, we flag the clusters that met the following criteria: (i) two clusters with a velocity offset $<$ 1000~\kms\ and (ii) physical separation is less than the sum of their $R_{500}$ values\footnote{Note that we recalculated the $R_{500}$ values using the redshift and $M_{500}$ values from the catalog for our adopted cosmology.}. We consider the most massive one in our analysis if two or more clusters satisfied these two conditions. This resulted in 186,378 `unique' clusters. This `unique' cluster sample spans a redshift range of  0.01$-$1.1 with a median redshift of 0.44.

\subsection{Quasar sample}  
\label{subsec:quasar-sample} 

We use the UV quasar catalog from the Hubble Spectroscopic Legacy Archive (HSLA) survey Data Release 2 \citep{2017cos..rept....4P}. This catalog contains spectra of 799 quasars observed with $HST$/COS. We select 583/799 quasars whose spectra are available in median resolution ($R \sim$~18,000) G130M and/or G160M gratings. The spectra obtained with G130M and G160M cover a useful spectral range between 1050$-$1450\AA\ and 1400$-$1800\AA\ respectively ,which are suitable for detecting the redshifted \Lya\ and UV metal absorption features. The SNR per resolution element of these quasars varies across a range of 4 to 19 with a median SNR of 8.

\subsection{Quasar-cluster pairs}  
\label{subsec:qso-cluster-pairs}

We cross-match the catalog of 583 background quasars with the 186,378 galaxy clusters with spectroscopic redshifts. We impose two selection criteria for our initial quasar-cluster pairs: (i) The line-of-sight (LOS) velocity offset between the quasar and cluster redshifts is $>5000$~\kms\ in order to minimize possible contamination due to absorption intrinsic to the background quasar and/or the quasar host-galaxy \citep[see e.g.,][]{Muzahid2013}. (ii) The projected separation between the foreground cluster and the background quasar at the redshift of the cluster (\rhocl) should be $< 10 R_{500}$, suitable to probe the cluster outskirts beyond several virial radii.

Our search after imposing the above-mentioned conditions yielded a total of 3230 quasar-cluster pairs with 2808 `unique' clusters and 431 `unique' quasars.\footnote{Many of the quasars are probing multiple clusters while some clusters are probed by more than one quasar.} In our subsequent analysis, following the method outlined in \cite{Mishra2021}, we found broad absorption line (BAL) features associated with the quasar emission in 7 quasar sight-lines, accounting for a total of 21 quasar-cluster pairs. In addition, one quasar sight-line accounting for 18  quasar-cluster pairs was unusable due to a Lyman Limit System (LLS) at z = 0.812. We exclude these 39 quasar-cluster pairs from our sample. Therefore, our final sample consists of 3191 quasar-cluster pairs with 2785 and 423 unique clusters and quasars, respectively.\par

In Fig.~\ref{fig:sample}, we show the scatter plot of cluster mass ($M_{500}$) vs redshift (\zcl) for the 2785 clusters ({\tt left} panel) and \nrhocl\ vs \zcl\ for the 3191 quasar-cluster pairs ({\tt right} panel). The cluster redshifts range from $0.01$ to $0.76$ with a median of $0.19$. The $M_{500}$ values range from $0.2-12.9 \times10^{14}$~\Msun\ with a median of $1.3\times10^{14}$~\Msun. The normalized cluster impact parameter (\nrhocl) of our quasar-cluster pairs ranges from $0.1 - 10.0$ by design, with a median value of $7.0$. Finally, the median cluster impact parameter of the quasar-cluster pairs in  our sample is 4.8 Mpc.\par

\subsection{Continuum normalization}

191 of the 423 quasars have both G130M and G160M grating spectra, while 198 (34) quasars are observed only with the G130M (G160M) grating. The G130M and G160M grating spectra of a given quasar are simply joined together from the wavelength at which the average SNR per pixel from the two gratings becomes roughly equal. Before continuum normalization, we exclude the spectral region from 1210$-$1220~\AA\ and 1301$-$1307~\AA\ to avoid the geocoronal \Lya\ and O\,{\sc i} emissions, respectively. In addition, we exclude the $\pm$100\kms\ region around the known strong galactic absorption lines. We bin each quasar spectrum by 3 pixels using the {\tt Python} routine {\it SpectRes} \citep{2017arXiv170505165C}. For continuum normalization of individual quasar spectra, we adopt a similar approach outlined in \citet{Mishra2022}.

Briefly, we first smooth each spectrum by 111 pixels to generate the pseudo-continuum level. This pseudo-continuum is then used to normalize each spectrum. We perform iterative boxed sigma-clipping with asymmetric sigma levels on the pseudo-continuum-normalized spectrum to ensure efficient clipping of absorption features while retaining the residual emission lines. The number of boxes is chosen based on the presence/absence of strong emission lines, while the asymmetric sigma levels are determined based on the median SNR of each spectrum. The iterative clipping is performed until the residual spectrum is sufficiently free from any absorption features. The spectrum is then interpolated linearly over the clipped spectral regions. We finally fit a spline over this interpolated spectrum. The knots of the spline are optimized based on the presence of prominent emission lines (such as the \Lyb, \OVI, \Lya, and \CIV) and the SNR of each spectrum. This is done to take into account the larger curvature near the emission lines while preventing over-fitting in noisy spectral regions.

\section{Results}
\label{sec:results}

\begin{figure*}
      \includegraphics[width=0.85\textwidth]{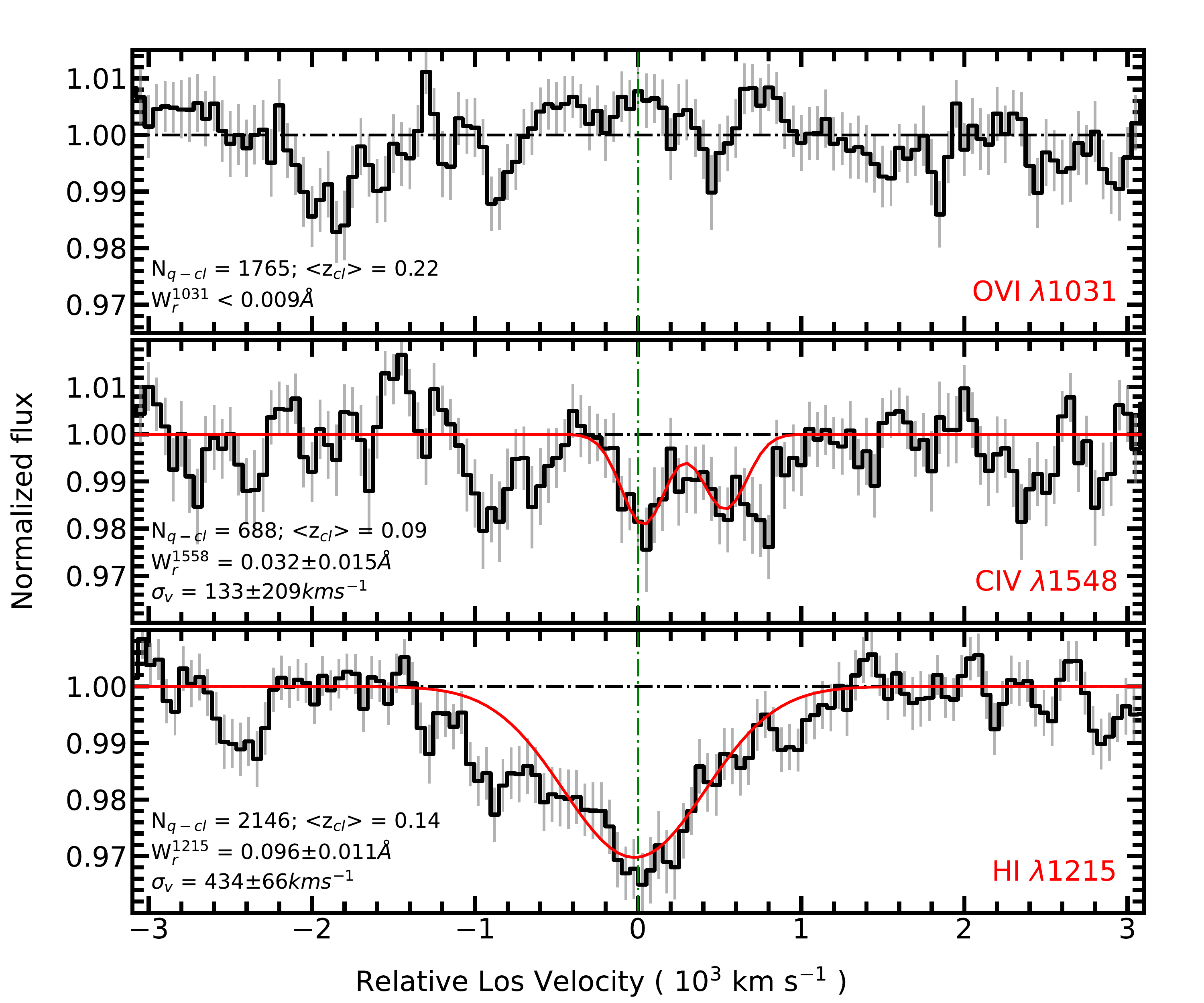}
     \caption{SNR-weighted mean stacked absorption profiles of \Lya\ ({\em bottom}), \CIV\ ({\em middle}), and \OVI\ ({\em top}) in the velocity scale at the rest-frame of the clusters. The grey $1 \sigma$ error-bars in each panel are estimated from 200 bootstrap realizations. Number of quasar-cluster pairs contributing to the stacked spectrum, median cluster redshift and rest-frame equivalent width (a 3$\sigma$ upper limit for \OVI) estimated from the stacked profile within $\pm$ 500\kms and $\pm$ 300\kms for \Lya\ and \CIV\ respectively around the line centroid is indicated in each panel. The best–fit single Gaussian component for the \Lya\ and \CIV\ doublet absorption profiles are shown in red and the corresponding velocity dispersion from the fits are also indicated.}   
     \label{fig:stackprofiles} 
\end{figure*}

\subsection{Spectral stacking of the full sample} 
\label{subsec:stacking}
\begin{table*}
\begin{adjustwidth}{-1cm}{}
\small
\caption{Summary of the measurements performed on the stacks of \Lya\ and metal lines.} 
\label{tab:results} 
\begin{tabular}{@{}ccccccccccc@{}} 
\hline \hline
Species  & $N_{\rm pairs}$ & \zcl\ & $M_{500}$ & $R_{500}$ & \rhocl  & \nrhocl & REW    &  $\sigma_{v}$  &  $V_{0}$   \\
    &                 &       & ($10^{14}~\mathrm{M}_{\odot}$)  & (Mpc)  & (Mpc) &  & (\AA) & (\kms) &  (\kms) \\ 
(1) & (2) & (3) & (4) & (5) & (6) & (7) & (8) & (9) & (10) \\
\hline
H\,{\sc i}~$\lambda$1215    &  2146        &  0.14(0.06$-$0.26)  &  1.26(0.78$-$2.38)  &  0.72(0.61$-$0.89)  &  4.93(2.67$-$6.93)  &  7.0(3.7$-$9.2)  &  0.096$\pm$0.011(0.040$\pm$0.005)  &  434$\pm$66   &  $-$22$\pm$62    \\
C\,{\sc iv}~$\lambda$1548   &  688         &  0.09(0.04$-$0.13)  &  1.19(0.75$-$2.37)  &  0.72(0.61$-$0.91)  &  5.12(2.77$-$7.20)  &  6.9(4.0$-$9.3)  &  0.032$\pm$0.015(0.019$\pm$0.009)  &  133$\pm$209  &  32$\pm$84    \\
O\,{\sc vi}~$\lambda$1031   &  1765        &  0.22(0.14$-$0.39)  &  1.35(0.83$-$2.41)  &  0.70(0.61$-$0.85)  &  4.88(2.70$-$6.74)  &  7.1(3.9$-$9.2)  &  $<$0.009         &  $-$          &  $-$          \\
C\,{\sc iii}~$\lambda$977    &   1236 & 0.29(0.20$-$0.45)  & 1.43(0.90$-$2.38)  &  0.70(0.61$-$0.84)  &  4.88(2.75$-$6.66)  &  7.1(4.0$-$9.2)  &  $<$0.010   &  $-$ & $-$ \\
C\,{\sc ii}~$\lambda$1036    &   1816 & 0.22(0.13$-$0.38)  & 1.34(0.83$-$2.38)  &  0.70(0.61$-$0.85)  &  4.89(2.67$-$6.75)  &  7.1(3.9$-$9.2)  &  $<$0.007   &  $-$ & $-$ \\
C\,{\sc ii}~$\lambda$1334    &   1537 & 0.12(0.04$-$0.23)  & 1.22(0.76$-$2.37)  &  0.72(0.61$-$0.89)  &  4.93(2.72$-$6.96)  &  6.9(3.8$-$9.2)  &  $<$0.016   &  $-$ & $-$ \\
Si\,{\sc ii}~$\lambda$989    &   1350 & 0.27(0.18$-$0.43)  & 1.42(0.89$-$2.40)  &  0.71(0.61$-$0.85)  &  4.87(2.74$-$6.66)  &  7.0(4.0$-$9.1)  &  $<$0.010   &  $-$ & $-$ \\
Si\,{\sc ii}~$\lambda$1193   &   2162 & 0.14(0.07$-$0.28)  & 1.26(0.78$-$2.37)  &  0.71(0.61$-$0.88)  &  4.92(2.70$-$6.87)  &  7.0(3.8$-$9.2)  &  $<$0.009   &  $-$ & $-$ \\
Si\,{\sc iii}~$\lambda$1206  &   2175 & 0.14(0.06$-$0.26)  & 1.26(0.78$-$2.38)  &  0.72(0.61$-$0.88)  &  4.93(2.69$-$6.92)  &  7.0(3.8$-$9.2)  &  $<$0.009   &  $-$ & $-$ \\
Si\,{\sc ii}~$\lambda$1260   &   1948 & 0.13(0.06$-$0.25)  & 1.26(0.77$-$2.43)  &  0.72(0.61$-$0.89)  &  4.95(2.69$-$6.96)  &  7.0(3.7$-$9.2)  &  $<$0.017   &  $-$ & $-$ \\
Si\,{\sc iv}~$\lambda$1393   &   1271 & 0.12(0.04$-$0.21)  & 1.22(0.78$-$2.37)  &  0.72(0.61$-$0.89)  &  4.94(2.70$-$6.96)  &  6.9(3.7$-$9.3)  &  $<$0.015   &  $-$ & $-$ \\

\hline 
\end{tabular} 
\end{adjustwidth}

\begin{tablenotes}\small
\item Notes -- (1) Name of the species around which stack profile is generated. (2) Number of quasar-cluster pairs. (3), (4), (5), (6), and (7) median values of cluster redshift, $M_{500}$, $R_{500}$, \rhocl, and \nrhocl~ respectively. The values in the parenthesis for the parameters listed from (3) $-$ (7) indicate the 16 and 84  percentiles of the parameter distribution. (8) REWs measured within $\pm$500\kms and $\pm$300\kms for \Lya\ and \CIV\ respectively and the upper limits on the REWs for other ions are estimated from the SNR-weighted mean stacked spectra. The values in the parenthesis are the REWs measured from the median stacked spectra. (9) \& (10) Velocity dispersion and Velocity centroids obtained from Gaussian fitting as explained in Section~\ref{subsec:stacking} in the mean stacks of \Lya\ and \CIV.
\end{tablenotes}
\end{table*}
To detect and analyze diffuse, cool/warm ($T\sim 10^{4-6}$~K) gas in the outskirts of the clusters, we employ the spectral stacking technique \citep[see][]{Mishra2022}. For each quasar-cluster pair, we first shift the quasar spectrum to the rest-frame of the cluster by dividing the observed wavelengths by ($1+z_{\rm cl}$). We only consider spectral regions between 1135$-$1450\AA\ and 1400$-$1790\AA\ for the G130M and G160M gratings, respectively, in the observed frame of each quasar. Whenever present, the spectral region blue-ward of the Lyman continuum break of a Lyman limit system is excluded. We only select those quasar-cluster pairs for our stacking analysis where the quasar spectra contribute at least 5 pixels with SNR per pixel $>1$ within a velocity window of $\pm$500\kms\ centered on a given line (i.e., \Lya, \OVI~$\lambda1031$, and \CIV~$\lambda1548$).

We populate the normalized raw flux values from the spectra of all the quasar-cluster pairs within a velocity range spanning $\pm 5000$~\kms\ in the rest from of the clusters. We use a bin size of $50$ km/s, which is about three times the spectral resolution of COS ($\Delta v \approx 18$~\kms). We confirm that our results are insensitive to the bin size used by varying the bin size to 100, 150, and 200~\kms. Subsequently, we compute the mean flux weighted by the median SNR\footnote{We only use spectral region within $\pm 5000$ km/s around the clusters to estimate the median SNR of the spectra.} of the spectra in each velocity bin. The SNR-weighted mean is preferred over the other statistics such as the median and mean, since the resultant stack profiles are not sensitive to the SNR of the individual spectra. In Section~\ref{appendix:snr_effect}, we demonstrate this with a comparative analysis of different stacking methods using mock quasar spectra.

The SNR-weighted mean composite spectra of \Lya, \OVI~$\lambda1031$, and \CIV~$\lambda1548$ are shown in Fig.~\ref{fig:stackprofiles}. The number of quasar-cluster pairs contributing to the \Lya, \OVI, and \CIV\ stacks are 2146 (1828 unique clusters and 404 unique quasars), 1765 (1686 unique clusters and 349 unique quasars), and 688 (594 unique clusters and 201 unique quasars), respectively. The stacked profiles shown in Fig.~\ref{fig:stackprofiles} are normalized by the corresponding pseudo-continua. To define a robust pseudo-continuum for each profile, we fit the first-order polynomials to randomly selected line-free regions between [$-5000, -1000$]~\kms\ and [$+1000, +5000$]~\kms\ for 1000 times. In each iteration, we apply $\sigma$-clipping with a threshold randomly chosen between 1 and 10~$\sigma$ levels. The mean of these 1000 realizations of the pseudo-continua yields the final pseudo-continuum. To estimate the rest-frame equivalent width (REW) of a stacked absorption, we subtract the contribution from the corresponding pseudo-continuum. As can be seen from the {\tt bottom} panel of Fig.~\ref{fig:stackprofiles}, the \Lya\ line is detected at with  99\% confidence level (CL) in the stacked spectrum with REW, \ewlya~$=0.096\pm0.011$~\AA. The uncertainty in the \ewlya\ is estimated by quadratically adding the statistical uncertainty calculated from the stacks of 200 bootstrap realizations\footnote{We confirm that increasing the bootstrap realizations to 500 or 1000 does not alter the results or conclusions drawn in this study.} of the 2146 quasar$-$cluster pairs and the uncertainty in the pseudo-continuum placement. We fit a single component Gaussian to the stacked \Lya\ absorption. The best-fit Gaussian has a velocity centroid of $-22\pm 62$~\kms\ and a velocity dispersion (\siglya) of $434 \pm 66$~\kms. The errors in the velocity centroid and velocity dispersion values are determined from the standard deviations of these values obtained from the fitting of the absorption profiles corresponding to the 200 bootstrap realizations. The median SNR achieved in the line-free region of the \Lya\ stack is $\approx 320$, corresponding to a 3$\sigma$ detection sensitivity of 0.009~\AA, assuming a spread of 21 pixels corresponding to the full-width at half maximum (FWHM) of the \Lya\ line ($\approx$ 2.355$\times \sigma_{\scaleto{\rm v, \HI \rm}{5pt}} \approx 1020$~\kms.

Next, we detect marginal \CIV\ absorption with CL of more than 95\% and with \ewciv\ of $0.032\pm0.015$~\AA, despite the fact that the number of quasar$-$cluster pairs contributing to the \CIV\ stack profile is significantly smaller (i.e., 688). We fit the \CIV\ doublet with a single Gaussian component. The velocity width of the two doublet Gaussians are kept the same but the amplitudes are allowed to vary. This yield a velocity centroid and velocity dispersion of $32\pm 84$~\kms\ and $133 \pm 209$~\kms, respectively. The fit suggests that the \CIV\ absorption is partially saturated. While the velocity centroid is consistent with $0$~\kms, the velocity dispersion is $\approx3$ times narrower than that of the \Lya, suggesting that the metal-bearing gas correlates over smaller velocity scale compared to the \Lya. The median SNR attained within the line-free region of the \CIV\ stack is 180 giving a 3$\sigma$ detection sensitivity of 0.020~\AA.

Finally, no significant \OVI\ absorption is detected in the high SNR composite spectrum shown in the top panel of Fig.~\ref{fig:stackprofiles}. From the \OVI\ stack with SNR of 256, we obtained a $3\sigma$ upper limit on the \OVI\ REW (\ewovi) of $ 0.009$~\AA.

We also put  constraints on the upper limits of the REWs of other ions such as \CII, \CIII, \SiII, \SiIII, and \SiIV, based on the non-detection of signals in the composite spectra. Table~\ref{tab:results} summarizes the results obtained from the stacking analysis.

\begin{figure*}
\begin{center}
     \includegraphics[width=0.9\textwidth]{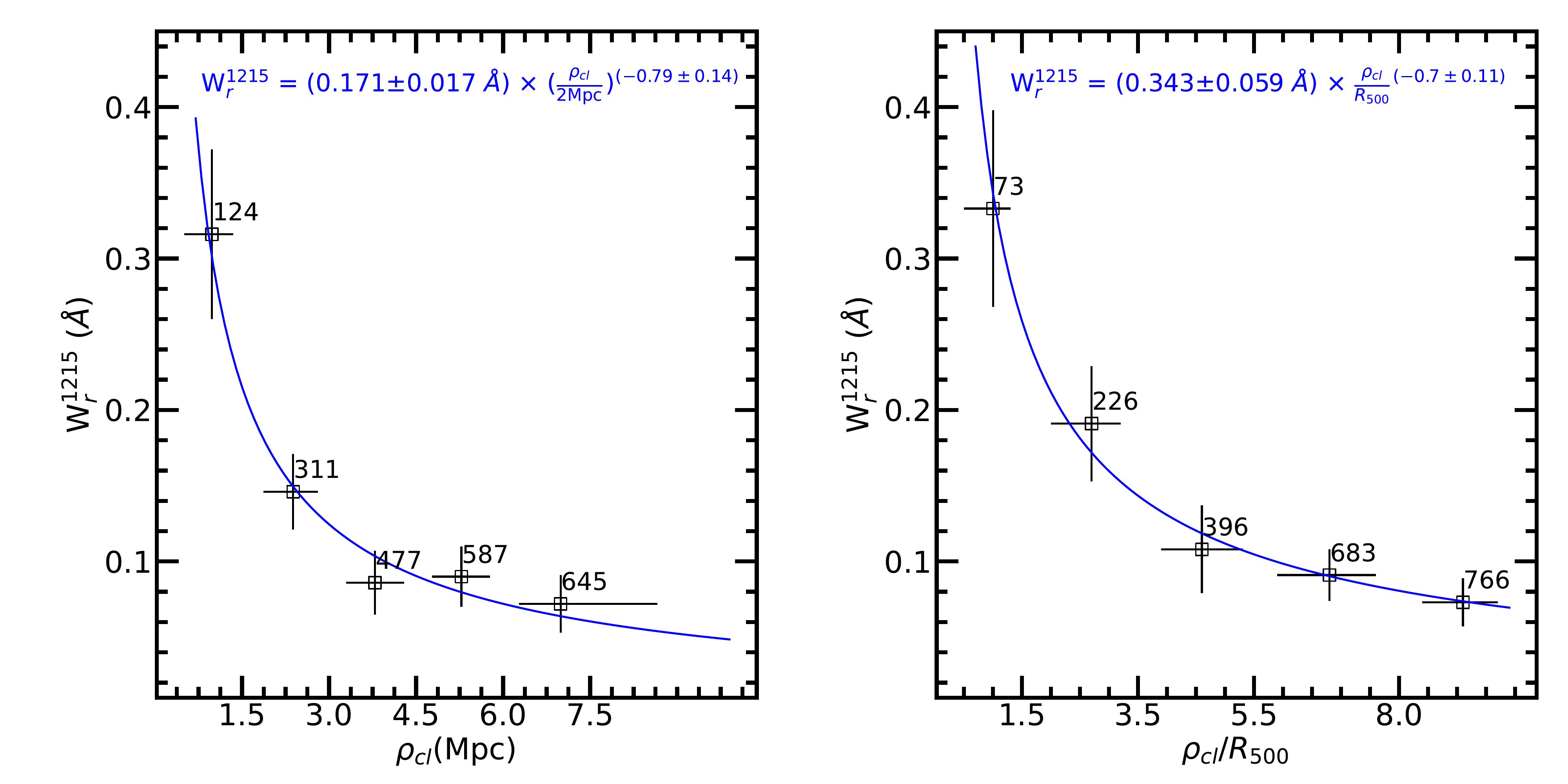}
     \caption{{\em Left:} REW of \Lya\ absorption as a function of \rhocl. The x-axis error bars represent the 68 percentile range of \rhocl\ in each bin. The y-axis error bars give the 1$\sigma$ scatter of \ewlya\ estimated from adding quadratically the statistical uncertainty of 200 bootstrap realizations and continuum placement uncertainty. The number of quasar$-$cluster pairs that contribute to each \rhocl\ bin is labeled in black next to each data point. The best-fitting power-law relation between \ewlya\ and \rhocl\ is indicated at the top side and is plotted in solid blue line. {\em Right:} similar to left but for \nrhocl.}
     
     \label{fig:result_bins}
\end{center}
\end{figure*}

\subsection{The \Lya\ equivalent width--profiles}   
\label{sub:rewprof}

To construct the \Lya\ REW-profile, we split the sample of 2146 quasar$-$cluster pairs contributing to the \Lya\ stack into five bins of \rhocl\ and \nrhocl\ using binsize of $1.5$Mpc and 2 respectively. Table~\ref{tab:results_subsample} summarises the details of the measurements performed on the \Lya\ subsample stacks. Fig.~\ref{fig:result_bins} shows the \Lya\ REW-profiles as a function of \rhocl\ ({\tt left} panel) and \nrhocl\ ({\tt right} panel). A declining trend in \ewlya\ with increasing \rhocl\ and \nrhocl\ is evident in Fig.~\ref{fig:result_bins}. As can be seen from Fig.~\ref{fig:result_bins}, a single power-law can explain the REW-profiles adequately with power-law slope of $-0.74\pm0.17$ for the \ewlya--\rhocl\ profile ({\tt left} panel) and $-0.60\pm0.15$ for the \ewlya--\nrhocl\ profile ({\tt right} panel). We do not find any correlation between the line width (\siglya) of the \Lya\ absorption signal with the \rhocl\  or \nrhocl.\par

Besides, we investigate the dependence of \ewlya\ on redshift and cluster mass by dividing the sample into three redshift and mass bins while maintaining a similar distribution of \nrhocl\ in all three bins of redshift and mass. We find no significant dependence of \ewlya\ on redshift. However, we find a tentative positive correlation between \ewlya\ and cluster mass, but the data points in the three mass bins are consistent within 1$\sigma$. Given the small number of quasar$-$cluster pairs for \CIV\ stack and the absence of significant signal in the \OVI\ stack, we could not perform similar analysis for the \CIV\ and \OVI\ stacks.

\subsection{Covering fraction analysis}
\label{subsec:covering_fraction}

In this section we determine the covering fractions (CFs) of the commonly detected UV transitions such as the \Lya, \CIV, and \OVI\ in the outskirts of the galaxy clusters by visual inspection of the relevant parts of the quasar spectra. The CF of a given transition is defined as:  
\begin{equation}
 f_c^{s} = \frac{ N_{\rm det}({W}^{s}_{r} \ge {W_{\rm th}}) } { N_{\rm tot}({W}^{s}_{\rm lim} \le {W_{\rm th}}) }~. 
\label{eqn:covfrac} 
\end{equation}
Here, $N_{\rm det}$ is the number of quasar-cluster pairs for which the given line is detected with a REW ($W^{s}_{r}$) more than a threshold equivalent width ($W_{\rm th}$), and $N_{\rm tot}$ is the total number of quasar$-$cluster pairs for which the quasar spectra are sensitive to detect the $W_{\rm th}$ (i.e., the $5\sigma$ limiting equivalent width, $W^{s}_{\rm lim}$, is lower than the $W_{\rm th}$). For each transition, we calculated the $W^{s}_{\rm lim}$ over a line-free region within $\pm$500\kms\ of the cluster redshifts using equation~6 of \citet{Hellsten1998}. \par

\begin{figure*}
\begin{center}
     \includegraphics[width=1.0\textwidth]{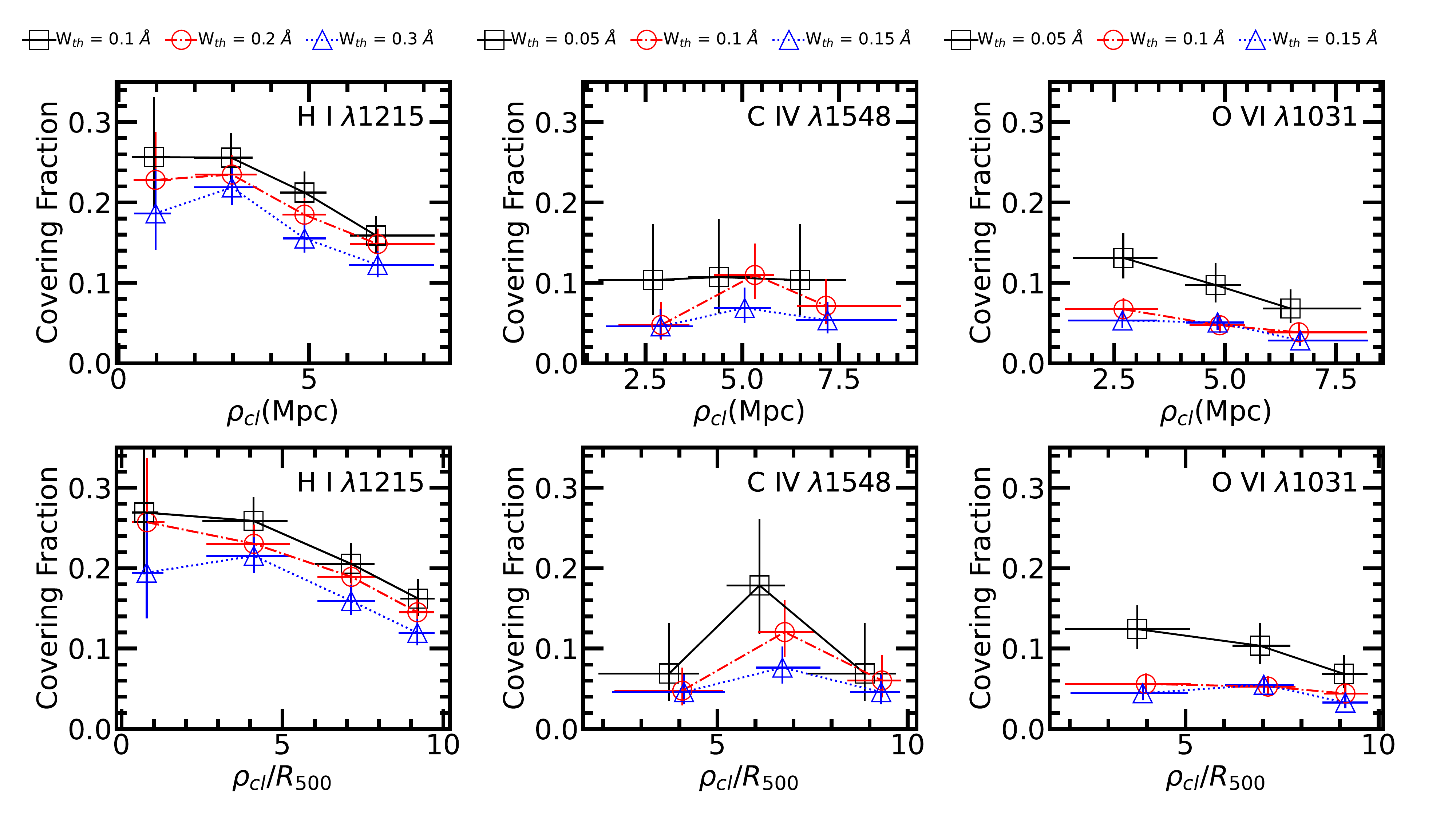}
     \caption{{\em Left:} Covering fraction of \Lya\ absorption systems of Flag-1 and -2 as a function of \rhocl\ ({\em top}) and \nrhocl\ ({\em bottom}) for three threshold values of equivalent widths (W$_{\rm th}$). The x-axis error bars represent the 68 per cent confidence interval. The y-axis error bars represent the 1$\sigma$ Wilson score confidence interval. {\em Middle:} Same as the left panels but for \CIV\ absorption systems. {\em Right:} Same as left panels but for \OVI\ absorption systems.}
     \label{fig:covering_fraction1}
\end{center}
\end{figure*}

For visual identification of \Lya\ absorption systems, we start with the 2146 quasar--cluster pairs used for \Lya\ stacking (see Table~\ref{tab:results}). To confirm the presence of \Lya, we only select the pairs for which at least \Lyb, among the higher order Lyman-series lines, is covered. We point out that the requirement of the presence of \Lyb\ and/or higher-order lines, will miss out the weak, stand-alone \Lya\ absorbers for which the higher order lines, including \Lyb, is too weak to be detected. Nonetheless, as identifying all/most of the absorption lines in so many quasar spectra is beyond the scope of this paper, we adopt this approach to avoid false positives. We do not impose any constraints on the detection significance for the \Lyb\ or higher-order Lyman-series lines. This reduces our sample of 2146 quasar$-$cluster pairs to 1160 pairs in the foreground of 295 quasars. We only search a $\pm$500\kms spectral region\footnote{This corresponds to $\pm$1$\sigma_{v}$ for a typical Virgo-like cluster with a median $M_{500} \sim$10$^{14.1}$~\Msun.} around the cluster redshifts for identifying \Lya\ absorption lines. We flag our detected \Lya\ absorption systems in two categories (Flag-1 and Flag-2) based on the confidence level of the detection. Flag-1 represents the high confidence systems where the \Lya\ absorption is accompanied by at least one higher-order Lyman series line and at least one metal line (e.g., \CII, \CIII, \CIV, \SiII, \SiIII, \SiIV, and \OVI). The shape of the apparent column density profiles \citep[ACD; using equation 8 from][]{Savage1991} of the Lyman-series lines are also checked for consistency for this class. For Flag-2, the \Lya\ systems satisfy one of the following two conditions: (1) have only associated \Lyb\ absorption with consistent column density profiles (2) have one of the metal lines present, but the apparent column density profiles are not fully consistent with each other, indicating blends. In total, we identify 241 \Lya\ systems, with 128 Flag-1 and 113 Flag-2 absorbers (in 218 unique clusters towards 118 unique quasars). Taking into account both Flag-1 and -2 systems (only Flag-1 systems), the  CF of \Lya\ is $0.21_{-0.01}^{+0.01}$ ($0.11_{-0.01}^{+0.01}$) for $W_{\rm th} =0.1$~\AA, at the median \nrhocl~$\approx$ 7.1. \par

Among the 1160 quasar$-$cluster pairs probed by 295 quasars, we find that 128 clusters probed by 58 quasars have redshifts consistent within $\pm$500~\kms\ of each other, but at different impact parameters. To avoid association of an absorption system with multiple clusters at varying impact parameters, we remove these 128 clusters and estimate the CF of the \Lya\ for the remaining 1032 quasar$-$cluster pairs. The CF of this sample is $0.21_{-0.01}^{+0.02}$ ($0.12_{-0.01}^{+0.01}$) for Flag-1 and -2 systems (only Flag-1 systems) for $W_{\rm th} =0.1$~\AA\ which is consistent with the CF estimates for the full sample of 1160 quasar$-$cluster pairs.

Out of the 688 quasar$-$cluster pairs used for \CIV\ stacking, we only consider the 649 pairs for visual inspection of \CIV\ absorption systems, for which the quasar spectra cover the \civab\ lines simultaneously. Similar to \Lya, we only use the regions within $\pm$500~\kms\ around the cluster redshifts. We also assign two flags to the identified \CIV\ absorption systems based on the  detection confidence. Flag-1 is assigned to the \CIV\ absorption systems that have two lines of the \CIV\ doublet with consistent ACD profiles, in addition to at least one Lyman series line or one other metal line. Systems containing only the two lines of the \CIV\ doublet with consistent ACD profiles or the two lines of the \CIV\ doublet with slight mismatching ACD profiles in addition to the presence of other lines are marked as Flag-2. We identify 49 \CIV\ systems with 38 Flag-1 and 11 Flag-2 categories in the outskirts of 45 unique clusters (34 unique quasars). The CF of \CIV\, including Flag-1 and -2 systems (only Flag-1 systems) with $W_{\rm th} =0.05$~\AA\ is $0.10_{-0.04}^{+0.03}$ ($0.09_{-0.03}^{+0.04}$) at the median \nrhocl~$\approx 6.1$.  \par

Of the 649 quasar$-$cluster pairs examined for \CIV, we find 47 clusters close in redshift within $\pm$500~\kms\ of each other towards 22 quasars but at different impact parameters. Similar to \Lya, after excluding these 47 clusters, the \CIV\ CF of the remaining 602 quasar$-$cluster pairs with Flag-1 and -2 absorption systems is $0.12_{-0.04}^{+0.03}$ for $W_{\rm th} =0.05$~\AA\ which is fully consistent with the CF of the full sample of 649 quasar$-$cluster pairs.

Likewise, for the visual identification of \OVI\ absorption systems, from among the 1765 quasar$-$cluster pairs used in \OVI\ stacking, we only choose 1553 pairs that have simultaneous spectral coverage of the \oviab\ lines within $\pm$500~\kms\ of the clusters' redshifts. The identified \OVI\ systems are classified into Flag-1 and Flag-2 categories using the same conventions used for \CIV. We identify 114 \ovi\ systems (78 Flag-1 and 36 Flag-2) in the outskirts of 114 unique clusters towards 82 unique quasars. The CF of \OVI\ with $W_{\rm th} =0.05$~\AA\ is $0.10_{-0.02}^{+0.01}$ ($0.07_{-0.02}^{+0.01}$) for Flag-1 and Flag-2 systems (only Flag-1 systems) at a median \nrhocl~$\approx 6.9$.  \par

Of the 1553 quasar-cluster pairs searched for \OVI, 120 clusters probed by 52 quasars have comparable redshifts within $\pm$500~\kms, but different impact parameters. Similar, to \Lya\ and \CIV, after excluding these 120 clusters, the CF of the \OVI\ for the remaining 1433 quasar$-$cluster pairs is $0.11_{-0.02}^{+0.01}$ for Flag-1 and -2 systems with W$_{\rm th} =0.05$~\AA\, which is in agreement with the CF of the full sample of 1553 quasar$-$cluster pairs.\par

Table~\ref{tab:results_CF} summarizes the results of the covering fraction analysis. In Fig.~\ref{fig:covering_fraction1}, the covering fraction profiles (CF-profiles) as functions of \rhocl\ ({\tt top} panel) and \nrhocl\ ({\tt bottom} panel) for \Lya, \CIV, and \OVI\ absorption systems are shown in the {\tt left}, {\tt middle}, and {\tt right} panels, respectively, for three different threshold rest equivalent widths ($W_{\rm th}$). The CF-profiles in Fig.~\ref{fig:covering_fraction1} are based on the samples that include both Flag-1 and Flag-2 absorption systems. The corresponding CF-profiles only for the Flag-1 systems are shown in Fig.~\ref{fig:covering_fraction2}. As evident from  Fig.~\ref{fig:covering_fraction1}, we find a moderate decreasing trend in CF with increasing \rhocl\ and \nrhocl\ for the \Lya\ and \OVI\, while no trend is seen for the \CIV\ CF-profile. The covering fraction decreases with decreasing the sensitivity limit (i.e. increasing the $W_{\rm th}$). 
As discussed in Section~\ref{subsec:cf_profile} and \ref{subsec:radial_profile_metals}, the \Lya, \CIV, and \OVI\ covering fractions measured in the outskirts of low-$z$ galaxy clusters are significantly lower compared to the measurements in the CGM of galaxies.

\subsection{Probing the CGM of cluster galaxies}

\label{subsec:res_environment_effect}

\begin{figure*}
    \includegraphics[width=1.0\textwidth]{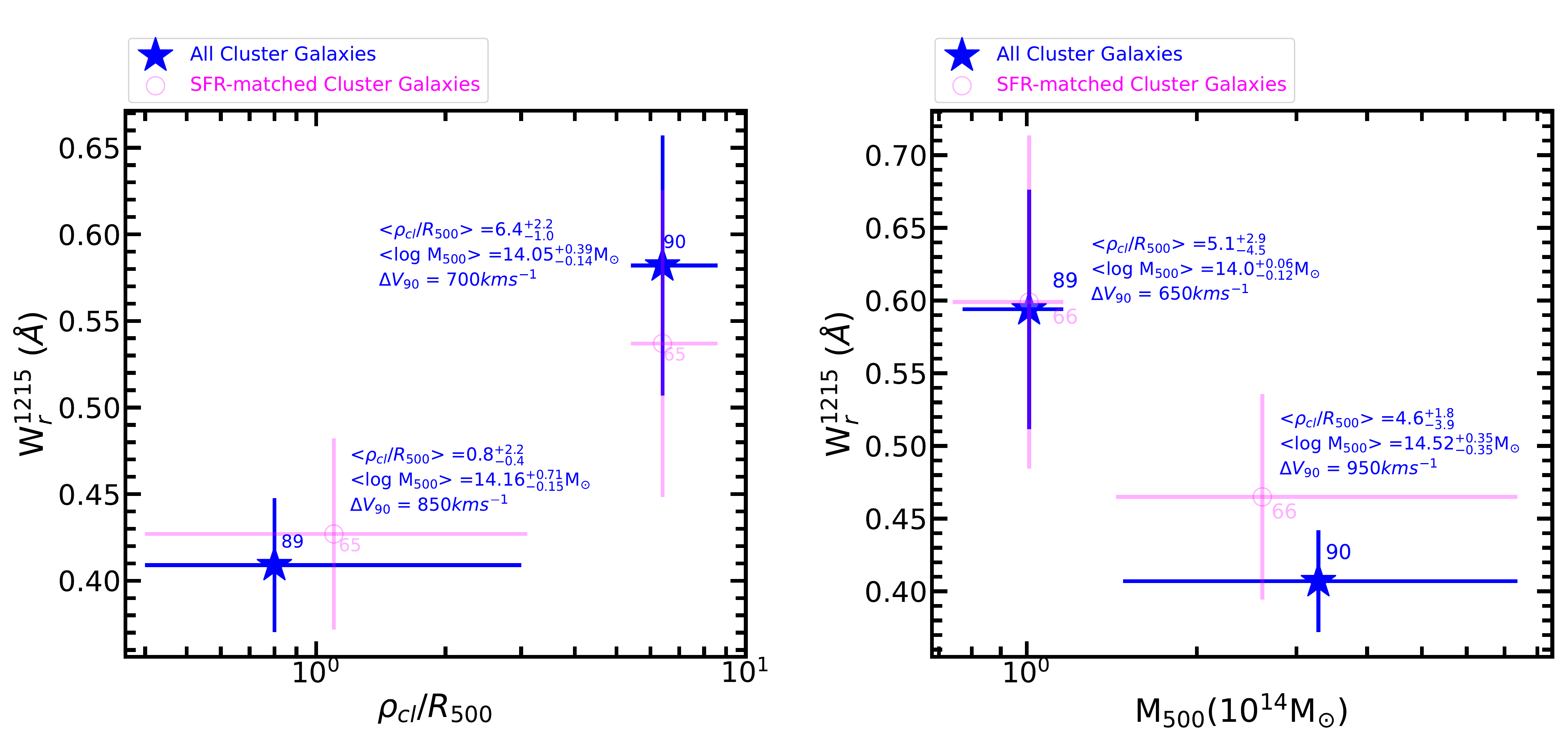}
     \caption{\emph{Left:} REW of \Lya\ in the rest-frame of the cluster galaxies (blue stars) as a function of cluster normalized impact parameter (\nrhocl). The median values of cluster mass, normalized clustocentric impact parameter, and $\Delta v_{90}$ of the stacked profile are indicated near the each data point. The REW of \Lya\ as a function of \nrhocl\ for the SFR-matched cluster galaxies are shown in open magenta circles. The number of galaxies contributing to each bin of \nrhocl\ for both samples are also indicated near the data points in respective colors. \emph{Right:} Similar to the left but as a function of $M_{500}$.}
     \label{fig:cluster_effect_on_CGM}
\end{figure*}

\begin{table*}
\begin{adjustwidth}{-1cm}{}
\small
\caption{Details of the measurements performed on the cluster galaxies.}
\label{tab:results_CLCGM} 
\begin{tabular}{ccccccccccccccccccccc}
    \hline

\multicolumn{1}{c}{Sample}       &  N  & \nrhocl  & $M_{500}$  & $z_{\rm gal}$  & $\log~(M_{\rm \star}/ M_{\odot})$  & $\log~[\rm SFR /M_{\odot}\ yr^{-1}]$  & $\rho_{\rm gal}$   & REW     & $\Delta v_{90}$ \\
                              &     &          & ($10^{14}~\mathrm{M}_{\odot}$) &      &   &  & (kpc)    & (\AA) & (\kms) \\
        (1)                   & (2) &   (3)    & (4)        &  (5)           &  (6)                               &  (7)                                &  (8)               &  (9)    & (10) \\

\hline 
         & \multicolumn{9}{c}{  \nrhocl\ Bin}  \\
\hline

\multirow{2}{*}{All Cluster Galaxies} & 89 & 0.8$_{-0.4}^{+2.2}$ & 1.43$_{-0.42}^{+5.94}$ & 0.03$_{-0.01}^{+0.03}$ & 9.8   $_{-0.6}^{+0.8}$ & -1.02$_{-0.90}^{+0.95}$ & 177$_{-70}^{+92}$ & 0.409$\pm$0.039 (0.170$\pm$0.033) & 850 \\
                                      & 90 & 6.4$_{-1.0}^{+2.2}$ & 1.13$_{-0.31}^{+1.63}$ & 0.04$_{-0.02}^{+0.04}$ & 10.0  $_{-0.9}^{+0.7}$ & -0.83$_{-0.68}^{+0.90}$ & 202$_{-88}^{+64}$ & 0.582$\pm$0.075 (0.288$\pm$0.030) & 700 \\\\

\multirow{2}{*}{SFR-matched Galaxies} & 65 & 1.1$_{-0.7}^{+2.0}$ & 1.36$_{-0.36}^{+6.01}$ & 0.03$_{-0.01}^{+0.04}$ & 9.7   $_{-0.5}^{+1.0}$ & -0.79$_{-0.88}^{+0.73}$ & 190$_{-91}^{+82}$ & 0.427$\pm$0.055 (0.167$\pm$0.028) & 900 \\
                                      & 65 & 6.4$_{-1.0}^{+2.2}$ & 1.12$_{-0.31}^{+1.64}$ & 0.04$_{-0.02}^{+0.03}$ & 10.1  $_{-0.9}^{+0.7}$ & -0.80$_{-0.84}^{+0.72}$ & 205$_{-73}^{+61}$ & 0.537$\pm$0.089 (0.232$\pm$0.036) & 649 \\\\

\hline

& \multicolumn{9}{c}{ $M_{500}$ Bin}  \\
\hline

\multirow{2}{*}{All Cluster Galaxies} & 89 & 5.1$_{-4.5}^{+2.9}$ & 1.01$_{-0.24}^{+0.15}$ & 0.04$_{-0.01}^{+0.04}$ & 10.0 $_{-0.9}^{+0.7}$ & -0.82$_{-0.69}^{+0.76}$ & 176$_{-73}^{+88}$ & 0.594$\pm$0.082 (0.314$\pm$0.029) & 650 \\
                                      & 90 & 4.6$_{-3.9}^{+2.8}$ & 3.28$_{-1.80}^{+4.09}$ & 0.02$_{-0.00}^{+0.05}$ & 9.9   $_{-0.7}^{+0.8}$ & -1.09$_{-0.83}^{+1.10}$ & 205$_{-89}^{+65}$ & 0.407$\pm$0.035 (0.173$\pm$0.020) & 950 \\ \\

\multirow{2}{*}{SFR-matched Galaxies} & 66 & 5.4$_{-4.7}^{+2.7}$ & 1.01$_{-0.27}^{+0.15}$ & 0.04$_{-0.01}^{+0.04}$ & 10.0 $_{-0.9}^{+0.5}$ & -0.88$_{-0.90}^{+0.81}$ & 175$_{-72}^{+89}$ & 0.599$\pm$0.115 (0.306$\pm$0.058) & 600 \\
                                      & 66 & 4.8$_{-3.2}^{+2.5}$ & 2.61$_{-1.17}^{+4.76}$ & 0.03$_{-0.01}^{+0.04}$ & 9.8  $_{-0.7}^{+0.8}$ & -0.91$_{-0.84}^{+0.88}$ & 208$_{-104}^{+72}$& 0.465$\pm$0.071 (0.188$\pm$0.043) & 750 \\

    \hline
\end{tabular}

\begin{tablenotes}\small
\item Notes -- (1) Galaxy sample. (2) Number of galaxies contributing to the stacks. (3), (4), (5), (6), (7), and (8) median values of \nrhocl, $M_{500}$, galaxy redshift, galaxy stellar mass, SFR, and galactocentric impact parameter. The uncertainties in columns (3) to (8) indicate the 68 percentile range. (9) REWs measured from the SNR-weighted mean spectra in the rest-frame of the galaxies within $\pm$500\kms. The values is the parenthesis are the REWs measured from the median stacked spectra. (10) The velocity difference between the pixels where optical depth is 5\% and 95\% of the total optical depth measured from the median stacked spectra.
\end{tablenotes}
\end{adjustwidth}
\end{table*}

To explore the influence of cluster environments on the CGM, for each cluster, we search for galaxies within impact parameters ($\rho_{\rm gal}$) of $< 300$~kpc from quasars and with spectroscopic redshifts within $\pm 1000$~\kms\ ($\Delta v$) of the cluster redshift. Note that these galaxies in the outskirts of clusters may or may not be member galaxies, but it is foreseen that they will eventually be part of the cluster. Hereafter, we will call them `cluster galaxies' for simplicity.

To identify the cluster galaxies, we use the $galSpec$ catalog\footnote{\url{https://www.sdss4.org/dr12/spectro/galaxy_mpajhu/}} by the Max Planck Institute for Astrophysics -- Johns Hopkins University (MPA-JHU) team. This catalog provides information on the star formation rate (SFR) and stellar mass of approximately 1.8 million galaxies with redshift $<0.33$ from SDSS DR8. The details of the stellar mass and SFR determination are explained in \citet{Brinchmann2004} and \citet{Kauffmann2003a}. We use the CasJobs SDSS SkyServer\footnote{\url{https://skyserver.sdss.org/casjobs/}} to obtain the galaxy properties from the $galSpec$ catalog. We only select galaxies with reliable spectroscopic properties with ``$\rm RELIABLE != 0$'' flags. We only consider galaxies for which both stellar mass and SFR information are available in the $galSpec$ catalog. For 94 clusters in the foreground of 83 unique quasars, we find 179 galaxies with $\rho_{\rm gal} <$ 300~kpc and |$\Delta$v| < 1000~\kms. We refer to this sample of 179 cluster galaxies with spectroscopic redshifts as the {\it CLCGM1000} sample. The median redshift and stellar mass of the galaxies in the {\it CLCGM1000} sample are $z_{\rm gal} \approx 0.04$  and $\log~(M_{\rm \star}/ M_{\odot}) \approx 9.9$ respectively. Similarly the median star formation rate (SFR) and specific SFR (sSFR) are $\log ~\rm [SFR /M_{\odot}\ yr^{-1}]\approx -0.9$ and $\log ~\rm [sSFR/yr^{-1}]\approx -10.76$, respectively.\par

Next, we divide the {\it CLCGM1000} sample into two bins of normalized clustocentric impact parameter (\nrhocl) and two bins of cluster mass ($M_{500}$) based on the corresponding median values. The median cluster masses of the two \nrhocl-bins are consistent with each other within $1\sigma$. The median \nrhocl\ values of the two mass-bins are also consistent within $1\sigma$. We generate SNR-weighted mean stacked \Lya\ profiles in the rest-frame of the galaxies contributing to each bins of \nrhocl\ and $M_{500}$. The \Lya\ REWs obtained from the stacking as a function of \nrhocl\ and  $M_{500}$ are shown with blue stars in the {\tt left} and {\tt right} panels of Fig.~\ref{fig:cluster_effect_on_CGM}, respectively.  Table~\ref{tab:results_CLCGM} presents the summary of the measurements obtained from this analysis. It is evident from the {\tt left} panel of Fig.~\ref{fig:cluster_effect_on_CGM} that the CGM of galaxies is relatively gas-poor when they are closer to clusters (\nrhocl~$< 4$). In addition, the {\tt right} panel shows that the CGM of galaxies is relatively gas-poor when they are in the outskirts of massive clusters ($M_{\rm 500} > 1.2\times10^{14}$~\Msun). The measured $\Delta V_{90}$ \footnote{The velocity difference between the pixels where optical depth is 5\% and 95\% of the total optical depth.} values suggest that the differences in REWs in the comparing bins are not owing to the significantly different line widths but due to the difference in optical depths. If at all, the $\Delta V_{90}$ values are somewhat higher for the bins showing lower REWs.
The broader $\Delta V_{90}$ values exhibited by the CGM of galaxies closer to  cluster centres or those residing in more massive clusters further suggests that the  gas clouds within the CGM of these galaxies is subject to higher velocity dispersion likely owing to various environmental effects.
\par

Even though the median stellar mass and SFR of the galaxies are consistent within 0.3 dex for the two bins of \nrhocl\ and $M_{500}$, a two-sample Kolmogorov–Smirnov (KS) test suggests a marginal difference in the SFR distribution of the galaxies in the two bins of \nrhocl\ and $M_{500}$.{\footnote{With the null probability that the two distributions are drawn from the same parent population being $\lesssim5$\%.}} A strong correlation between \Lya\ REW and SFR within the virial radius of (non-cluster) galaxies is recently reported by \citet{SDutta2023}. We thus generated SFR-matched subsamples with a tolerance of 0.1~dex for the above exercises. The measurements corresponding to the SFR-matched subsamples are shown by the open magenta circles in Fig.~\ref{fig:cluster_effect_on_CGM}. The figure shows that even after matching the SFR of the galaxies, the trends persist. Consequently, it can be inferred that the trends are governed by external environments rather than internal galactic processes.

It is evident from Fig.~\ref{fig:cluster_effect_on_CGM} that the CGM of cluster galaxies that are closer or reside in massive clusters is considerably deficient in cool, neutral gas as compared to the CGM of galaxies farther from the cluster centers or towards low-mass clusters. These observations indicate that the CGM is strongly influenced by the large-scale environments of galaxies. In a follow-up paper, we are planning to present the covering fractions of \HI\ and other metal lines for these galaxies.

\section{Discussion}
\label{sec:discussion}

Here we report on the first detection of cool, neutral gas traced by \lya\ along with a marginal detection of \CIV\ in the SNR-weighted mean stacked spectra of $2146$ and $688$ quasar$-$cluster pairs, respectively, at a median \nrhocl~$\approx 7$. We obtain an upper limit of $\sim 0.009$\AA\ ($3\sigma$) for \OVI\ based on the non-detection in the high-SNR composite spectrum of $1765$ quasar$-$cluster pairs. Furthermore, we determine the CF of $0.21_{-0.01}^{+0.01}$ for $W_{\rm th} =0.1$~\AA\ for \Lya, $0.10_{-0.04}^{+0.03}$ for $W_{\rm th}  =0.05$~\AA\ for \CIV, and $0.10_{-0.02}^{+0.01}$ for $W_{\rm th} =0.05$~\AA\ for \OVI\ in the cluster outskirts (median \nrhocl~$\approx7$). We also find that the CGM of cluster galaxies exhibits a significant environmental dependence. In this section, we discuss the distribution and origin of the cool, neutral gas and metals detected in the cluster outskirts, and the role of environment in determining the gas reservoirs in the CGM of galaxies.

\subsection{Distribution of neutral gas in cluster outskirts: \Lya\ REW--profile}
\label{subsec:radial_profile}

The distribution of cool, neutral gas as a function of impact parameter for low-mass haloes ($M_{\rm vir} \lesssim \rm 10^{13}$~\Msun) has been studied extensively in the literature \citep[e.g.,][]{Tumlinson2013,Wilde2021,SDutta2023}, with several studies indicating a decline in the \Lya\ REW as the impact parameter from the hosting haloes increases. At high mass scales ($> \rm 10^{14}$~\Msun) of galaxy clusters, on the other hand, observations and simulations of the hot ICM show that the temperature, density, and metal content of hot gas vary significantly from the core to the outskirts  \citep{Nagai2011,Simionescu2011,Urban2017}, which may influence the distribution of cool, neutral gas surrounding clusters. For instance there are studies suggesting that the cool gas traced by the \Lya\ absorbers is shock heated to high temperatures in the ICM, avoiding the inner region of the clusters \citep[see][]{yoon2012}. However, there are observations of cool gas in the hot ICM. For example, the detection of extended H$\alpha$ emission surrounding the galaxy NGC 1275, located at the center of the Perseus cluster \citep[see][]{Conselice2001}. The H$\alpha$ emitting gas may have originated from the cooling of the ICM, as suggested by \citet{Fabian1994}. Additionally, simulations of galaxy clusters revealed that fast, radiatively cooling outflows can also cause condensation of cool gas \citep[e.g., see][]{Qiu2021}. Therefore, the REW--profile of \Lya\ as a function of cluster impact parameter, which has not been explored thoroughly using statistically significant samples of clusters, is crucial.

Using 23 background quasar sightlines, \citet{yoon2012} found that \Lya\ absorbers avoid the central X-ray emitting regions of the Virgo cluster. Moreover, relatively stronger \Lya\ absorbers (REW~$>0.3$~\AA) are found to be associated with Virgo substructures. Excluding the \Lya\ absorbers associated with the substructures, they reported a weakly increasing \Lya\ REW with an increasing clustocentric impact parameter, which is in contrast with CGM observations. A similar analysis for the Coma, using 8 background quasars, also revealed no significant trend between \Lya\ REW and impact parameter \citep[see][]{Yoon2017}. Similarly, \citet{Muzahid2017} found no significant trend between the \HI\ column density and normalized cluster impact parameter up to $5 R_{500}$ (see their figure 3) by combining their results with those of \citet{yoon2012}, \citet{Yoon2017}, and \citet{Burchett2018}.

In this study, the \Lya\ REW--profile, constructed using a statistically significant sample of 2146 quasar-cluster pairs, shows a significant decline with 
 the impact parameter (and normalized impact parameter) as reported in the literature for galactic halos. A single-component power-law can adequately explain the REW--profiles (see Fig.~\ref{fig:covering_fraction1}).  \footnote{Suggested by an F-test with $>95$\%  confidence.} The best-fitting relations are: 
 \begin{equation} 
W_r^{1215} = (0.171 \pm 0.017~ \text{\AA}) \times (\frac{\rho_{\rm cl}}{{\rm 2~Mpc}})^{-0.79\pm0.14}, 
\end{equation}
 and 
\begin{equation} 
W_r^{1215} = (0.343 \pm 0.059~ \text{\AA}) \times (\frac{\rho_{\rm cl}}{R_{500}})^{-0.70\pm0.11}~.  
 \end{equation}

Recently, \citet[]{SDutta2023} showed that the \Lya\ REW-profile for low-$z$ galaxies can be decomposed into two components: (i) a Gaussian/log-linear component, prominent only at small scales ($\lesssim R_{200}$), and (ii) a power-law component, reminiscent of galaxy-absorber clustering, describing the large-scale features. Owing to the smaller sample size, particularly at \rhocl~$<R_{500}$ ($R_{200}$), and much weaker overall absorption signal compared to the CGM of galaxies (see next section for details), we could not investigate the possible existence of similar characteristics in the \Lya\ REW--profile around low-$z$ clusters.

\subsection{Distribution of neutral gas in cluster outskirts: \Lya\ covering fraction} 
\label{subsec:cf_profile}

We determine the covering fraction of \Lya\ using a subsample of quasar-cluster pairs with simultaneous coverage of the \Lya\ and \Lyb\ lines. We obtained $\rm CF = 0.21^{+0.01}_{-0.01}$ ($0.16^{+0.01}_{-0.01}$) for a threshold REW of $0.1$~\AA\ ($0.3$~\AA) for the full sample. The \Lya\ CF--profiles obtained for the full sample, shown on the {\tt left} panels of Fig.~\ref{fig:covering_fraction1}, show a declining trend with increasing \rhocl\ and \nrhocl\ (see also Fig.~\ref{fig:covering_fraction2}). Overall, the \Lya\ CF remains in the range of 10--25\% which is broadly consistent with the predictions from hydrodynamical simulation \citep[see figure 3 of][]{Butsky2019} over a length scale of 3 Mpc from clusters. The \Lya\ covering fraction increases rapidly to $\approx 1$ in the simulation for \rhocl~$<500$~kpc ($\approx 0.7 R_{\rm 200}$). We are yet to probe this impact parameter range observationally.

When a LOS velocity of $\approx 1000$~\kms\ is used to associate \Lya\ absorbers with Coma, the CF is measured to be $\approx 33$\% for $W_{\rm th}> 0.1$~\AA\ ($\approx25$\% for $W_{\rm th} > 0.3$~\AA) both within and outside $R_{\rm vir}$ \citep[see][]{Yoon2017}. For the less massive Virgo cluster, the authors reported a marginally ($\approx1\sigma$) higher \Lya\ CF outside the $R_{\rm vir}$ ($0.38^{+0.14}_{-0.12}$ for $W_{\rm th} > 0.1$~\AA). The covering fraction measurements for Virgo and Coma presented by \citet[]{Yoon2017} have large uncertainties, and are broadly consistent with our measurements. Here we caution that our CF measurements could be systematically smaller as we required the presence of \Lyb\ (or higher order Lyman series lines) for detection, which will miss the stand-alone, weak \Lya\ absorbers.

Next we compare the CF of \Lya\ absorbers in the outskirts of galaxy clusters with the CF measured in the CGM of galaxies. Several studies in the literature have constrained the \Lya\ CF in the CGM of galaxies across a wide range of redshift, stellar mass, SFR, and impact parameter \citep[see][]{Tumlinson2013, Liang2014, Keeney2018, Wilde2021}. For instance, \citet{Tumlinson2013} found CF of 100\% and 75\% for star-forming and passive galaxies, respectively, for \Lya\ absorbers with $N(\HI) > 10^{13}$~\sqcm\ within $0.5 R_{\rm vir}$. Meanwhile, the CF of \Lya\ absorbers with $W_{\rm th} >0.05$~\AA\ in \citet{Liang2014} remained above 60\% out to $8R_{\rm vir}$. \citet{Keeney2018} reported 50\% CF for $N(\HI) > 10^{13}$~\sqcm\ \Lya\ absorbers in both $< L*$ and $\ge$ $L*$ galaxies up to $4 R_{\rm vir}$. Recently, \citet{Wilde2021} showed that the CF of \Lya\ absorption with $N(\HI) = 10^{13-15}$~\sqcm\ is nearly 80\% over 4$R_{\rm vir}$ from galaxies. Overall, most CGM studies have reported a high CF of  $>$50\% for \Lya\ absorbers out to $4R_{\rm vir}$. In contrast, our study finds a low CF of \Lya\ absorption systems with W$_{\rm th} >$ 0.1\AA\ (i.e., $N(\HI) > 10^{13.3}$~\sqcm) ranging from  $\approx25$\% at $<$ $R_{500}$ to $\approx15$\% at around $6 - 10 R_{500}$ (see Fig.~\ref{fig:covering_fraction1}). 

\par

\subsection{Distribution of metals in the cluster outskirts}
\label{subsec:radial_profile_metals}

We produce stacks of all the prominent metal lines accessible to COS. However, except for a marginal \CIV, none of the metal lines are detected (see Table~\ref{tab:results}). By examining the relevant parts of the COS spectra, we determine the covering fraction of the \CIV\ and \OVI, the two most commonly studied metal lines. For \CIV, we obtain $\rm CF = 0.10^{+0.03}_{-0.04}$ for $W_{\rm th} = 0.05$~\AA. The CF for \OVI\ is found to be similar (i.e., $0.10^{+0.01}_{-0.02}$ for $W_{\rm th}= 0.05$~\AA).

The covering fraction of \CIV\ and/or \OVI\ are not well constrained in the earlier studies for the cluster outskirts. Using a sample of only 7 cluster pairs, \citet{Tejos2016} obtained a CF of $0.14_{-0.1}^{+0.3}$ for \OVI\ with $W_{\rm th} = 0.06$~\AA. The CF value is consistent with our measurement. \citet{Burchett2018} found no \OVI\ absorption within $\pm1500$~\kms\ of the 5 clusters in their study. This is also consistent with the low \OVI\ CF measured here.

The distribution of metals traced by \CIV\ and \OVI\ absorption have been explored in the simulation of \citet{Butsky2019}. They found  CF of \CIV\ within $\approx 2R_{200}$ to be $<0.1$, whereas for \OVI, it is $>0.15$. The authors attributed this difference to the lower ionization potential of \CIV\ (48 eV) compared to the \OVI\ (114 eV), which makes it challenging to sustain the triply-ionized state for carbon at temperatures of $>10^{5}$~K. However, for both the ions, the covering fraction increases significantly in the inner 500~kpc regions. The \CIV\ covering fraction we estimate in the impact parameter range of 500--3000~kpc is somewhat higher compared to the simulation prediction \citep[i.e., only a few percent; see figure 3 of][]{Butsky2019}. The \OVI\ CF of $\approx10$\%, however, is consistent with our measurements (see Fig.~\ref{fig:covering_fraction1}).

The non-detection of \OVI\ in the stack of 1765 quasar-cluster pairs is intriguing. It is generally believed that the gas temperature in cluster outskirts is more suitable for the tracers of the warm-hot phase, such as the \OVI. However, the non-detection of \OVI\ in the high SNR stacked spectra puts a stringent upper limit on the $N(\OVI)$ of $10^{12.7}$~\sqcm. Combined with the marginal detection of \CIV, we obtain an upper limit on the gas temperature of $10^{5.3}$~K, under collisional ionization equilibrium \citep[CIE;][]{Gnat2007} assuming single-phase gas with solar relative abundances. It indicates that cluster outskirts are not potential reservoirs of warm-hot gas.

We point out that the marginal detection of \CIV\ in the stack of 688 quasar-cluster pairs while the non-detection of \OVI, albeit with a $\approx2.5$ times larger sample size and a covering fraction similar to \CIV, could be due to two factors: (i) The REW of the detected \CIV\ absorption systems tends to be somewhat higher compared to \OVI. (ii) The LOS velocity distribution of the detected absorbers shows a somewhat larger scatter for \OVI.

It is noteworthy that among the 49 \CIV\ absorption systems, 44 systems show associated \Lya\ absorption (median $\rm REW = 0.8$~\AA), while the remaining 5 \CIV\ systems do not have spectral coverage for \Lya. Likewise, all 114 \OVI\ absorbers in our study are associated with \Lya\ (median $\rm REW = 0.7$~\AA) and/or \Lyb\ absorption. The association of the strong \Lya\ absorption with the metal lines suggests that a fraction of these systems may be arising from the CGM of cluster galaxies close to the sightline near a background quasars. In the next section we will estimate the contribution of the circumgalactic gas to the stacked signals.

\subsection{Contribution of the CGM to the stacked signals} 
\label{subsec:cluster-galaxy-asso}

\begin{figure}
      \includegraphics[width=0.48\textwidth]{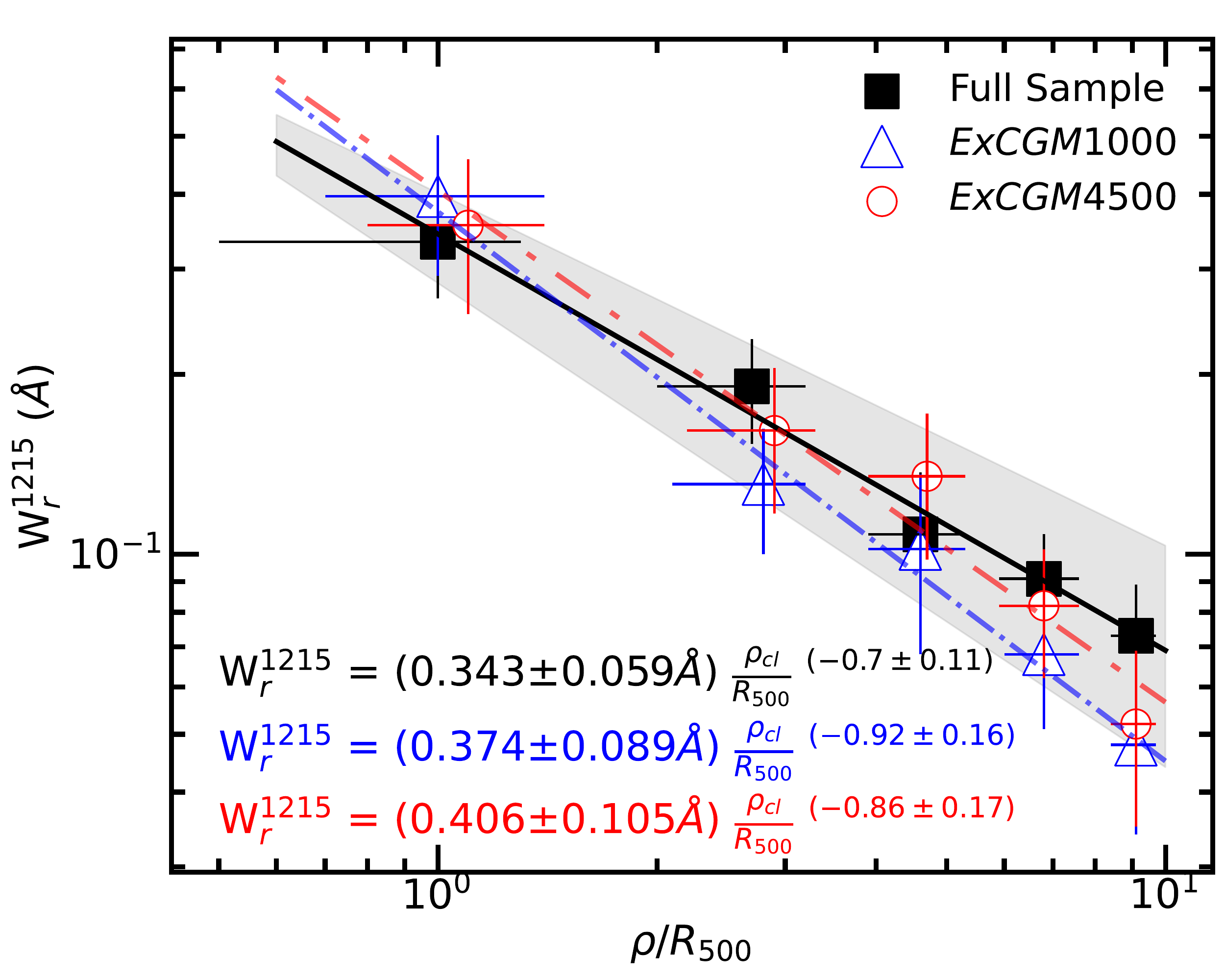}
     \caption{The \Lya\ REW profile for the full sample (black squares), {\it EX-CGM1000} sample (blue triangles) and {\it EX-CGM4500} sample (red circles) as a function of normalized impact parameter (see Section~\ref{subsec:cluster-galaxy-asso}). The grey shaded region represents the 1$\sigma$ scatter around the measurements of the full sample. The best fit single-component power law relations for these samples are also indicated.} 
     \label{fig:CCM_vs_CGM_comp}
\end{figure}


In order to understand whether the \Lya\ absorption signals in the stacks are dominated by the CGM of cluster member galaxies, we adopt an approach similar to \citet{Mishra2022}. Similar to Section~\ref{subsec:res_environment_effect}, for each cluster, we first identify galaxies within a radius of 300~kpc surrounding the background quasar with photometric and/or spectroscopic redshifts consistent with that of the cluster. Since the majority of our clusters are from DESI, we use the photometric redshift catalog of galaxies from the DESI Legacy Imaging Surveys \citep{Zou2019} to identify galaxies around the clusters. The typical uncertainty of the photo-$z$ estimates of the DESI galaxies is $0.017$; we therefore use a LOS velocity window of $\pm4500$~\kms\ ($\equiv c \times \Delta z/1+z$) for linking galaxies to a cluster. Next, we eliminate those clusters from our analysis when one or more galaxies are found.

For 1140 of the 2146 quasar--cluster pairs, we identify a total of $2251$ cluster galaxies with a median $r \approx 19$. \footnote{Only $614$ of the $2215$ galaxies have spectroscopic redshifts.} We exclude those $1140$ quasar--cluster pairs, for which the background quasars may probe the CGM of cluster galaxies, and call the remaining sample of 1006 ($2146-1140$) quasar--cluster pairs the {\it ExCGM4500} sample. Furthermore, since the LOS velocity of $\pm4500$~\kms corresponds to a large cosmological distance of $\approx \pm 60$~Mpc at the median redshift of 0.14, we use one more conservative window of $1000$~\kms\ and label the corresponding sample as {\it ExCGM1000}. We identify a total of 694 galaxies associated with 476 quasar$-$cluster pairs within the velocity window of $\pm1000$~\kms\ in {\it ExCGM1000} sample.

The \Lya\ REWs measured for the {\it ExCGM1000} and {\it ExCGM4500} subsamples are $0.028\pm0.006$~\AA\ and $0.027\pm0.008$~\AA, respectively. Both are significant at the 99\% CL, and are consistent with the REW measured for the full sample within the $\approx 1.5\sigma$ allowed uncertainties. As can be seen from Fig.~\ref{fig:CCM_vs_CGM_comp}, the \Lya\ REWs measured for the sub-samples after excluding the possible CGM contributions are consistent with the full sample within $1\sigma$, indicating that the CGM of bright cluster galaxies in the outskirts does not dominate the \Lya\ absorption signal.

\subsection{Cosmic filaments as the origin of the stacked signals}  
\label{subsec:infall_substructure}

\begin{figure*}

 \includegraphics[width=0.9\textwidth]{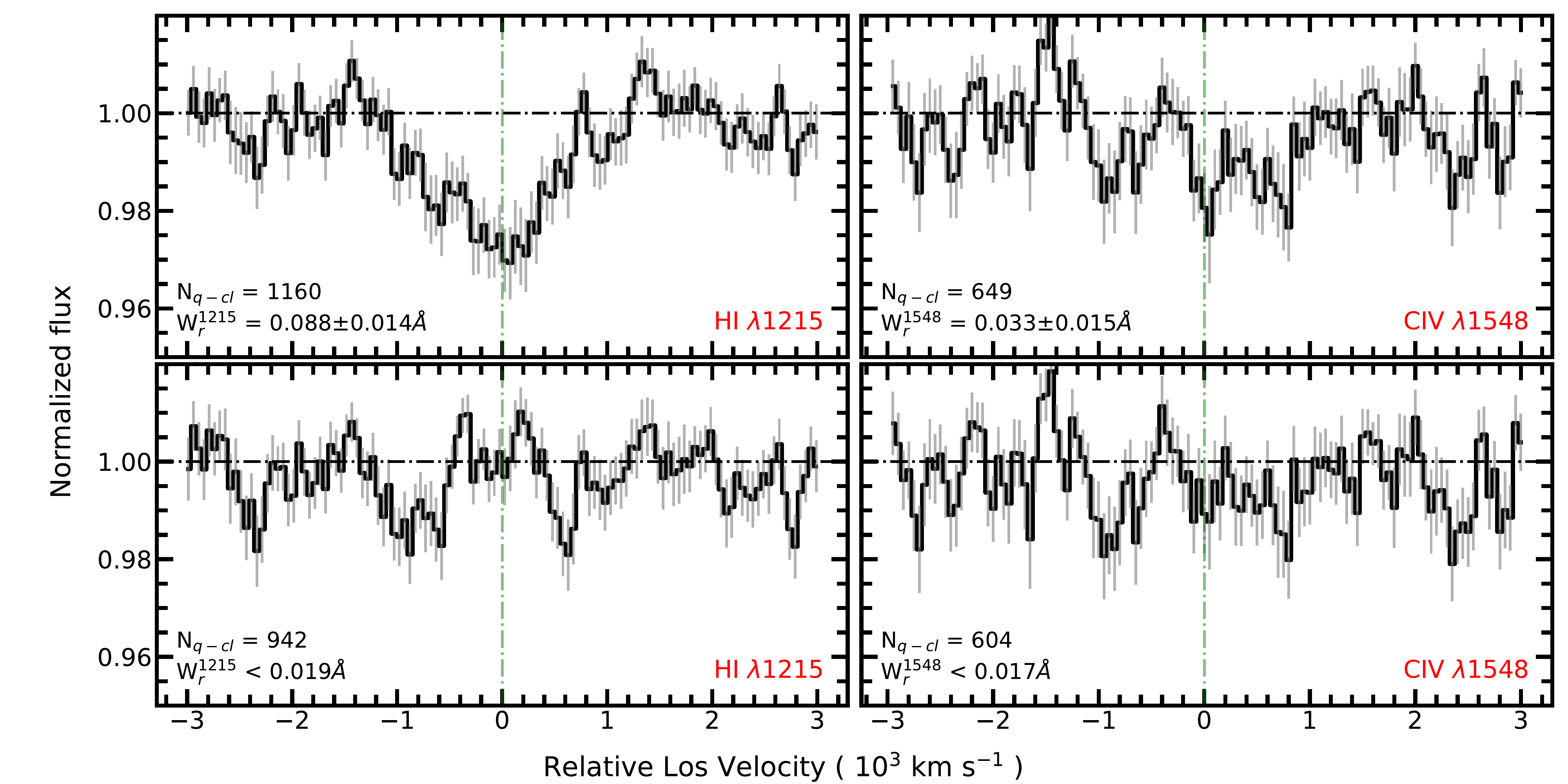}
 \caption{\emph{Left:} SNR-weigthed mean stacked \Lya\ profile for the sample of 1160 quasar$-$cluster pairs used in the covering fraction analysis (see Section~\ref{subsec:covering_fraction}) is shown in the top panel. The bottom left panel shows the stacked \Lya\ profile for the non-detection sample, i.e., pairs where no individual detection of absorption systems with Flag-1 or -2 is reported. The number of quasar$-$cluster pairs and REW (3$\sigma$ upper limit for non-detection) are shown at the lower left corner of each panel. \emph{Right:} The same as the left panel but for \CIV\ absorption.}
 \label{fig:composite_nondetecion} 
\end{figure*}

To investigate the origin of the observed cool, neutral gas and metals around clusters, we re-examine the samples of 1160 and 649 quasar$-$cluster pairs used for the covering fraction analysis of \Lya\ and \CIV\ (see Section~\ref{subsec:covering_fraction}). As mentioned earlier, we identify 241 and 49 \Lya\ and \CIV\ absorption systems, respectively, within $\pm$500\kms around 218 and 45 quasar$-$cluster pairs. We isolate the subsamples of 942 (= 1160 - 218) and 604 (= 649 - 45) quasar$-$cluster pairs without any individual absorption detection respectively for \Lya\ and \CIV\ and produce stacks. The {\tt top} panels of Fig.~\ref{fig:composite_nondetecion} show the stacked \Lya\ and \CIV\ profiles for the full samples of 1160 and 649 quasar-cluster pairs respectively. Notably, we detect \Lya\ and \CIV\ signals at with CL $>$99\% and 95\% CL, respectively. However, no significant absorption is seen for either \Lya\ or \CIV\ when we exclude the pairs with visually identified \Lya\ and \CIV\ absorbers (see the {\tt bottom} panels of Fig.~\ref{fig:composite_nondetecion}). To understand this non-detection of signal when excluding visually identified strong \Lya\ absorbers, we also conduct a mock stacking analysis of weak \Lya\ absorbers in Section~\ref{appendix:mock_weak_abs}. We find that the expected \Lya\ signal from the weak absorbers is less than the $3\sigma$ upper limit on the REW from the observation (see Fig.~\ref{fig:composite_nondetecion}). A detailed analysis of this finding is presented in Section~\ref{appendix:mock_weak_abs}.

This indicates that the signals in the full stacks ({\tt top} panels of Fig.~\ref{fig:composite_nondetecion}) are primarily dominated by the individual absorbers used in the covering fraction analysis. These findings indicate that the distribution of cool gas and metals in the cluster outskirts is patchy, and the detected absorption signals in the stacked spectra may arise from cosmic filaments feeding the clusters with cool gas rather than from a volume-filling gas phase. The observed low covering fractions of gas and metals, particularly in comparison with the CGM of galaxies, are also consistent with this picture. In line with our argument, \citet{Wakker2015}, in a study to probe nearby galaxy filaments using \Lya\ absorption, found a strong anti-correlation between \Lya\ REW and filament impact parameter, with absorbers with $N(\HI) > 10^{13}$~\sqcm\ primarily detected within $\approx500$~kpc from the filament axis \citep[see also][]{Bouma2021}.

We also search for the galaxies around the 241 \Lya\ absorption systems as identified in Section~\ref{subsec:covering_fraction}. We use the similar initial DESI galaxy catalog used for creating $ExCGM$ samples in  Section~\ref{subsec:cluster-galaxy-asso}. Around each \Lya\ system, we look for galaxies with photometric and/or spectroscopic redshifts within $|\Delta v| <$ 500, 1000, 2000, and 4000~\kms of the absorbers and a projected separation of $<$ 300~kpc. Out of 241 \Lya\ systems, we find 13\% (32 / 241), 28\% (68 / 241), 39\% (94 / 241), and 56\% (136 / 241) systems have associated bright galaxies (median $r \approx 19$) within $|\Delta v| <$ 500, 1000, 2000, and 4000~\kms, respectively. This suggests that most of the detected \Lya\ absorbers are likely unrelated to the halos of {\em bright galaxies}, and may instead originate from the cosmic filaments. However, conducting deeper galaxy surveys around the cluster outskirts is crucial to search for the presence of faint, low-mass galaxies around these absorbers. Alternatively, the observed absorption signal could arise from stripped off gas due to multiple environmental processes faraway from any galaxies as discussed below.

\subsection{Stripping as the origin of the stacked signals}  
\label{subsec:environment_effect}

Fig.~\ref{fig:cluster_effect_on_CGM} shows that the CGM of galaxies located in the proximity of clusters (i.e., \nrhocl~$< 4$) is more gas-deficient compared to their counterparts located farther away. Similarly, at comparable distances from clusters, the CGM of galaxies residing in more massive halos (i.e., $M_{\rm 500} > 1.2\times10^{14}$~\Msun) is more gas-deprived compared to galaxies in the outskirts of low-mass clusters. Since the galaxies in the comparable bins are matched in galaxy properties (e.g., SFR, $M_\star$), we argued that this difference in the gas content is not governed by internal galactic processes but by environmental effects. The dearth of cool, neutral gas in the CGM of galaxies residing close to cluster centres and/or residing in the outskirts of massive clusters can be understood in terms of efficient ram pressure stripping of circumgalactic matter that is loosely bound to the galaxies. Both the density of the ambient medium ($\rho_{\rm amb}$) and the relative velocity of infalling galaxies with respect to the ambient medium ($v_{\rm rel}$) are expected to be higher for massive clusters and/or in the inner regions of clusters. This naturally facilitates higher ram pressure ($P_{\rm ram} = \rho_{\rm amb} \times v_{\rm rel}^2$) stripping for galaxies residing in massive clusters and/or closer to cluster centers. Indeed, the recent simulation by \citet{rohr2023} revealed that the majority of the jellyfish galaxies lose  the bulk of their cold gas ($<10^{4.5}$~K) via ram pressure stripping within 1--2 Gyr since infall when they are at a distance as far as $3-4 R_{200}$. The authors further showed that satellite galaxies of all types deposit more than $10^{10}$~\Msun\ of cold gas via ram pressure stripping in hosts of mass $> 10^{13}$~\Msun\ over a time-scale of 5 Gyr. Such stripped gas faraway from galaxies can explain the observed absorption signals if the gas remains cold.

The other important environmental process that can cause gas loss closer to cluster centres ($<2-3 R_{200}$) is ``overshooting'' \citep[e.g.,][]{Pimbblet2011,McGee2009,Bahe2013,Borrow2023}. It is found in simulation that a substantial fraction of galaxies in cluster outskirts are actually backsplash galaxies, which have undergone a significant gas loss due to severe tidal disruption and stripping caused by ram pressure during their first pericentre passage. The current position of such backsplash galaxies and the locations at which their gas has been stripped in the outskirts can be considerably different. Such ``overshot'' galaxies can give rise to the signals detected here. This process, nevertheless, is not efficient beyond a few virial radii \citep[see e.g.,][]{Bahe2013}.

The results in the {\tt right} panel of Fig.~\ref{fig:cluster_effect_on_CGM} are intriguing as they indicate that galaxies residing in more massive clusters exhibit a significant deficiency of circumgalactic gas compared to those in less massive clusters, even at distances as far as $\approx 5R_{500}$. The lack of cool gas in the CGM of galaxies residing in high-mass clusters is likely due to ``pre-processing''. Gas removal through ``pre-processing'' happens when a galaxy becomes a satellites of a group (substructure) in the outskirts of a massive cluster prior to infall \citep[e.g.,][]{Wetzel2013,Jaffe2016,Hough2023}. According to the hierarchical structure formation model, more substructures are expected around massive halos, which facilitates efficient ``pre-processing'' of the CGM of galaxies residing in the outskirts of massive clusters.

It is expected that all of these processes discussed above (ram pressure stripping, overshooting, and pre-processing) will contribute to the cool gas cross-section in the cluster outskirts to a varied degree, depending on the host mass and distance from the core. However, the overall low covering fraction of the neutral gas in cluster outskirts suggests that the stripped circumgalactic gas is either mixing with the ambient hotter medium or is heated up by AGN feedback or that the galaxies supplying cool gas via stripping are predominantly falling onto clusters through large-scale cosmic filaments.

\section{Conclusion}
\label{sec:conlusion}

By cross-matching the HSLA quasar catalog with seven cluster catalogs from the literature, we have built a sample of 3191 quasar-cluster pairs with a median cluster redshift and mass of $0.19$ and $1.3\times10^{14}$~\Msun, respectively. The quasars are probing the clusters with impact parameters in the range 0.7--10~Mpc with a median of 4.8~Mpc (median \nrhocl~$\approx7$).  Using spectral stacking, we present the first detection of cool, neutral gas (traced by \Lya) and metals (traced by \CIV) in the outskirts of clusters. In addition, we construct the \Lya\ REW--profile out to  $10 R_{500}$. We visually identify 241 \Lya, 49 \CIV, and 114 \OVI\ absorption systems within $\pm$500~\kms\ of the cluster redshifts, which allow us to determine the covering fractions of \Lya\ and the other metal ions. We leverage the combined results of the stacking analysis and the distribution of individual absorption systems to investigate the nature and origin of the cool, neutral gas and metals in the outskirts of the clusters. The key results of this paper are summarized as follows:

\begin{itemize}
    \item In the SNR-weighted mean stacked spectra of $2146$ and $688$ quasar$-$cluster pairs, we detect significant \Lya\ and marginal \CIV\ absorption with REWs of $0.096\pm0.011$~\AA\ (at $>$ 99\% CL) and $0.032\pm0.015$~\AA\ (at $>$ 95\% CL) respectively. We do not detect \OVI\ in the high-SNR stacked spectrum of $1765$ quasar$-$cluster pairs giving a $3\sigma$ upper limit on the REW of $<0.009$~\AA\ (Fig.~\ref{fig:stackprofiles}).

    \smallskip
    \item The covering fractions of  \Lya-, \CIV-, and \OVI-bearing gas in cluster outskirts are $0.21_{-0.01}^{+0.01}$ (for $W_{\rm th} >$ 0.1~\AA), $0.10_{-0.04}^{+0.03}$ (for $W_{\rm th} >$ 0.05~\AA), and $0.10_{-0.02}^{+0.01}$ (for $W_{\rm th} >$ 0.05~\AA), respectively. These are significantly lower as compared to the measurements obtained for the CGM of galaxies at similar redshifts in the literature.

    \smallskip 
    \item  The REW-profile of \Lya\ exhibits a decreasing trend with increasing \rhocl\ (\nrhocl) which is well described by a power-law with a slope of $-0.79\pm0.14$ ($-0.70\pm0.11$; see Fig.~\ref{fig:result_bins}). Similarly, the covering fractions of \Lya\ and \OVI\ also gradually decrease with increasing \rhocl\ and \nrhocl, while no such trend is observed for the \CIV\ covering fraction profile (Fig.~\ref{fig:covering_fraction1}).

    \smallskip 
    \item We show that the \Lya\ REW-profiles for the $ExCGM1000$ and $ExCGM4500$ sub-samples are consistent with the REW-profile of the full sample within $1\sigma$ indicating that the signal is not dominated by the CGM of bright galaxies (Fig.~\ref{fig:CCM_vs_CGM_comp}).

    \smallskip 
    \item We do not find any detectable absorption in the stacked spectra when we remove the visually identified absorbers from our  analysis (Fig.~\ref{fig:composite_nondetecion}). Only 13\% of the visually identified \Lya\ absorbers are related to bright ($r\approx 19$) galaxies with $|\Delta v| < 500$~\kms\ and $\rho_{\rm gal} < 300$~kpc. Based on these, we argued that the distribution of the detected cool gas in the cluster outskirts is filamentary, which is also consistent with the observed low covering fractions.

    \smallskip 
    \item  We find that the CGM of galaxies that are closer to cluster centers or that are in massive clusters is significantly gas poor compared to the galaxies farther away from the clusters or residing in the low-mass clusters (Fig.~\ref{fig:cluster_effect_on_CGM}). We also find that the trends persist even after matching the SFR and $M_{\rm \star}$ of the cluster galaxies, indicating that the trends are governed by external environments rather than internal galactic processes.
\end{itemize}

We discuss the roles of environmental processes such as ram pressure stripping, overshooting, and pre-processing to understand the observed gas deficiencies in cluster galaxies. Our findings suggest that at smaller clustocentric distances, ram pressure stripping and overshooting are likely the dominant processes, whereas at large distances ($\gtrsim 4-5 R_{500}$) pre-processing dominates the stripping of circumgalactic gas. We argue that the cool gas detected in cluster outskirts arises in part from stripped-off CGM and from large-scale filaments feeding the clusters with cool gas. In the future, we plan to perform detailed photoionization modelling of the individual absorbers detected in the outskirts of clusters to constrain the physical conditions and metal enrichment and to determine the relative contributions of the two channels to the observed absorption signals.

\section*{ACKNOWLEDGEMENTS}

This research has made use of the HSLA database, developed and maintained at STScI, Baltimore, USA. We acknowledge the use of High performance computing facility PEGASUS at IUCAA.\par

The DESI Legacy Imaging Surveys consist of three individual and complementary projects: the Dark Energy Camera Legacy Survey (DECaLS), the Beijing-Arizona Sky Survey (BASS), and the Mayall z-band Legacy Survey (MzLS). DECaLS, BASS and MzLS together include data obtained, respectively, at the Blanco telescope, Cerro Tololo Inter-American Observatory, NSF’s NOIRLab; the Bok telescope, Steward Observatory, University of Arizona; and the Mayall telescope, Kitt Peak National Observatory, NOIRLab. NOIRLab is operated by the Association of Universities for Research in Astronomy (AURA) under a cooperative agreement with the National Science Foundation. Pipeline processing and analyses of the data were supported by NOIRLab and the Lawrence Berkeley National Laboratory (LBNL). Legacy Surveys also uses data products from the Near-Earth Object Wide-field Infrared Survey Explorer (NEOWISE), a project of the Jet Propulsion Laboratory/California Institute of Technology, funded by the National Aeronautics and Space Administration. Legacy Surveys was supported by: the Director, Office of Science, Office of High Energy Physics of the U.S. Department of Energy; the National Energy Research Scientific Computing Center, a DOE Office of Science User Facility; the U.S. National Science Foundation, Division of Astronomical Sciences; the National Astronomical Observatories of China, the Chinese Academy of Sciences and the Chinese National Natural Science Foundation. LBNL is managed by the Regents of the University of California under contract to the U.S. Department of Energy.

\section*{Data AVAILABILITY}
The spectral data underlying this paper are available in the HSLA Archive at \url{https://archive.stsci.edu/missions-and-data/hsla}. The DESI photo-z galaxy catalog is publicly available at \url{http://cdsarc.u-strasbg.fr/viz-bin/cat/J/ApJS/242/8}.

The cluster catalogs used in this study are also publicly available and can be obtained from the following links.\\ 
BL15: \url{https://cdsarc.cds.unistra.fr/viz-bin/cat/J/ApJS/216/27}; WH15: \url{https://cdsarc.cds.unistra.fr/viz-bin/cat/J/ApJ/807/178}; WHF18: \url{https://cdsarc.cds.unistra.fr/viz-bin/cat/J/MNRAS/475/343}; BL20: \url{https://cdsarc.cds.unistra.fr/viz-bin/cat/J/ApJS/247/25} H20: \url{https://iopscience.iop.org/article/10.3847/1538-3881/ab6a96\#ajab6a96app1}; H21: \url{https://cdsarc.cds.unistra.fr/viz-bin/cat/J/ApJS/253/3}; Z21: \url{https://cdsarc.cds.unistra.fr/viz-bin/cat/J/ApJS/253/56}.

\bibliography{references}

\setcounter{section}{0}
\setcounter{table}{0}
\setcounter{figure}{0}

\renewcommand{\thefigure}{A\arabic{figure}}
\renewcommand{\thetable}{A\arabic{table}}  
\renewcommand{\thesection}{A\arabic{section}}

\begin{center}
{\Large \sc Appendix}\\
\end{center}

\section{Brief notes on the cluster catalogs}
\label{appendix:catalog_summary}

We briefly describe below the seven cluster catalogs used in this study:

\citet{Bleem2015} compiled a sample of 677 Sunyaev-Zel'dovich (SZ) selected galaxy clusters from South Pole Telescope (SPT) in 2500 deg$^{2}$. The authors used a spatial-spectral matched filter and a simple peak-finding algorithm in the 95 and 150 GHz survey data to identify the clusters above the SNR threshold ($\xi$) of $~$4.5. Among these 677 clusters, the spectroscopic redshift for 169 clusters was computed using the follow-up observations from various telescopes (such as the VLT/FORS2 and the Gemini-South telescope/GMOS-S) and the literature wherever available. The purity of this sample is 95\% for the $\xi >$ 5. Therefore, we only consider clusters from this sample with spectroscopic redshift and $\xi >$ 5 yielding a total of 134 SZ clusters from this catalog. These clusters cover a redshift range of 0.06$-$1.5 and $M_{500}$ range of 2.4$-$17.5 $\times10^{14}$\(M_\odot\).\par


Using photometric redshifts of galaxies from the SDSS, \citet{Wen2015} identified a total of 158,103 clusters. The found 25,419 new clusters, and also updated the \citet{Wen2012} catalog of 132,684 clusters with spectroscopic redshifts, $R_{500}$, and richness ($R_{L*,500}\equiv L_{500}/L_{*}$), using scaling relations obtained from a sample of 1191 clusters with well-measured masses via X-ray or SZ-observations from the literature. We limit our cluster sample to the same mass and redshift ranges as the sample of 1191 clusters from which the mass scaling relation was derived. This resulted in a total of 156,139 clusters in the \citet{Wen2015} catalog. Additionally, we only select clusters with spectroscopic redshift, which provided a total of 119,653 clusters. Although no formal completeness estimate for the \citet{Wen2015} sample was provided, the \citet{Wen2012} catalog, which constitutes the majority of the sample of \citet{Wen2015}, is over 95\% complete for masses greater than $10^{14} \mathrm{M}_{\odot}$ in the redshift range of $0.05 \leq z < 0.42$. The resulting sample of 119,653 clusters covers a redshift range of 0.05 to 0.7 and a mass range of 0.3 to 31.4 $\times10^{14}$\(M_\odot\).\par 

By combining the photometric data from the Two Micron All Sky Survey (2MASS), Wide-field Infrared Survey Explorer (WISE), and SuperCOSMOS, \citet{Wen2018} compiled 47,000 clusters in the sky area of 28000 deg$^{2}$ among which 26,125 clusters were newly identified, and lie outside the SDSS coverage. The clusters were identified as the overdense region around the brightest cluster galaxy (BCG) candidates selected from the magnitude and colour cuts. The photometric redshifts of the clusters were obtained from the artificial neural network (ANNz) approach \citep{2016ApJS..225....5B}. Monte Carlo simulations showed that the false detection rate for their cluster sample drops below the 5\% for SNR $>$ 5. We only consider the clusters with spectroscopic redshift and SNR $>$ 5 resulting a total of 4454 galaxy clusters from this catalog. The redshift and mass range of this catalog are 0.03$-$0.45 and 0.51$-$11.72$\times10^{14}$\(M_\odot\) respectively.\par

\citet{Bleem2020} compiled a sample of 470 SZ-selected galaxy clusters from the SPTpol-Extended Cluster Survey (SPT-ECS) in the 2770 deg$^{2}$ at 95 and 150 GHz. Using a similar spatial-spectral matching filter technique as in the \citet{Bleem2015}, the authors identified 266 cluster candidates with $\xi >$ 5 and 204 clusters with 4 $< \xi <$ 5. The masses of the clusters were derived based on the $\xi$-mass scaling relation, inferred from fitting the observed SZ cluster density at $\xi >$ 5 and redshift z $>$ 0.25. The authors employed the red-sequence Matched-Filter Probabilistic Percolation (redMaPPer) algorithm \citep{2014ApJ...785..104R} and constrained the photometric redshift of their sample. The spectroscopic redshifts for 63 clusters were drawn from the literature. The survey is $>$ 90\% complete for z $>$ 0.25 at $\xi >$ 5. Limiting only to the clusters with spectroscopic redshifts and with z $>$ 0.25 at $\xi >$ 5, we consider only 26 clusters from this catalog. These 26 clusters span a redshift range of 0.25 < z < 1.39 and the mass range of 3.1 < $M_{500}$ < 14.6$\times10^{14}$\(M_\odot\).\par

\citet{Huang2020} presented 89 SZ-selected clusters at 95 and 150 GHz in 100 deg$^{2}$ with the SPTpol receiver on the SPT with a SNR greater than 4.6. The authors used the follow-up and archival optical/infrared images and spectra to confirm the clusters and estimate photometric redshifts for 66 clusters and spectroscopic redshifts for 23 clusters. This catalog is $>$ 95\% complete for clusters with $M_{500c} >$ 2.6 $\times10^{14}$\(M_\odot\) h$^{-1}$ and z $>$ 0.25. Restricting only to clusters with spectroscopic redshift and with $M_{500c} >$ 2.6 $\times10^{14}$\(M_\odot\) h$^{-1}$ and z $>$ 0.25, we consider 11 clusters from this survey. These clusters cover a redshift range of 0.25 < z < 1.38 and the mass range of 2.6 < $M_{500}$ < 8.4$\times10^{14}$\(M_\odot\).\par

Using 98 and 150 GHz observations with the AdvACT receiver on the Atacama Cosmology Telescope (ACT) in 13,211 deg$^{2}$ of the sky, \citet{Hilton2021} identified 4195 SZ-selected and optically confirmed galaxy clusters with SNR $>$ 4. Assuming an SZ-signal vs mass scaling relation calibrated from X-ray observations, the sample has a 90\% completeness mass limit of $M_{500c} >$ 3.8 $\times10^{14}$\(M_\odot\) evaluated at z = 0.5, for SNR $>$ 5. The photometric redshift for 2547 clusters and the spectroscopic redshift for 1648 clusters were drawn from the literature. We restrict our search to those clusters with spectroscopic redshift and SNR $>$ 5, which resulted in a total of 1176 clusters. The redshift and mass range for these clusters are 0.04$-$1.91 and 1.2$-$13.5$\times10^{14}$\(M_\odot\), respectively.\par

Using a fast clustering algorithm, \citet{Zou2021} identified 540,432 z $\lessapprox$ 1 clusters with photometric redshift in the DESI legacy imaging surveys, covering a sky area of 20,000 deg$^{2}$. The mass of the clusters were derived using a calibrated richness–mass relation that is based on the observations of X-ray emission and the SZ effect. Monte Carlo simulations indicated that the false-detection rate of their cluster identification algorithm is about 3.1\%. In our study, we consider only 122,390 clusters with spectroscopic redshifts. The redshift and mass coverage of these 122,390 clusters are 0.01$-$1.61 and 0.1$-$10.9$\times10^{14}$\(M_\odot\), respectively.\par

\begin{table*}
\small
\caption{Summary of the measurements performed on the stacks of the subsamples of \Lya.} 
\label{tab:results_subsample} 
\begin{tabular}{@{}ccccccc@{}} 
\hline
Bin  & \rhocl\   & $N_{\rm pairs}$ &   REW & $\sigma_{v}$  &  $V_{0}$   \\
\rhocl(Mpc) &      Mpc        &      &  \AA   &  \kms          & \kms\        \\
(1) & (2) & (3) & (4) & (5) & (6)  \\
\hline

$<$1.5             &  0.98$^{+0.48}_{-0.37}$  & 124    &  0.316$\pm$0.061 (0.106$\pm$0.026)  &  444$\pm$65   &     $-$51 $\pm$105    \\  
$[$1.5$-$3.0$)$    &  2.38$^{+0.51}_{-0.43}$  & 311    &  0.146$\pm$0.033 (0.051$\pm$0.014)  &  391$\pm$75   &     167 $\pm$97       \\  
$[$3.0$-$4.5$)$    &  3.79$^{+0.50}_{-0.50}$  & 477    &  0.086$\pm$0.024 (0.041$\pm$0.011)  &  616$\pm$226  &     $-$185$\pm$205    \\  
$[$4.5$-$6.0$)$    &  5.28$^{+0.51}_{-0.49}$  & 587    &  0.090$\pm$0.022 (0.037$\pm$0.011)  &  494$\pm$145  &     $-$27 $\pm$151    \\  
$\ge$6.0           &  6.99$^{+0.72}_{-1.60}$  & 647    &  0.072$\pm$0.022 (0.022$\pm$0.008)  &  623$\pm$118  &     $-$217$\pm$235    \\\\
\hline

Bin &  \nrhocl\   &  $N_{\rm pairs}$ & REW  &  $\sigma_{v}$  &  $V_{0}$   \\
\nrhocl &              &      &  \AA   &  \kms          & \kms\        \\
(1) & (2) & (3) & (4) & (5) & (6)  \\
\hline

$<$1.5         &  1.0 $^{+0.50}_{-0.30}$    & 73    &  0.333$\pm$0.071 (0.113$\pm$0.029) & 377$\pm$99   &     $-$152$\pm$126    \\  
$[$1.5$-$3.5$)$&  2.7 $^{+0.70}_{-0.50}$    & 226   &  0.191$\pm$0.042 (0.067$\pm$0.019) & 457$\pm$58   &     169 $\pm$103      \\  
$[$3.5$-$5.5$)$&  4.6 $^{+0.70}_{-0.70}$    & 396   &  0.108$\pm$0.030 (0.041$\pm$0.013) & 548$\pm$207  &     $-$213$\pm$291    \\  
$[$5.5$-$8.0$)$&  6.8 $^{+0.90}_{-0.80}$    & 683   &  0.091$\pm$0.021 (0.035$\pm$0.008) & 444$\pm$172  &     110 $\pm$68       \\  
$\ge$8.0       &  9.1 $^{+0.70}_{-0.59}$    & 768   &  0.073$\pm$0.018 (0.029$\pm$0.008) & 522$\pm$137  &     $-$297$\pm$154      \\

\hline 
\end{tabular} 

\begin{tablenotes}\small
\item Notes -- (1) Bin size. (2) Median values of \rhocl\ (\nrhocl) for \rhocl-bins (\nrhocl-bins). (3). Number of quasar$-$cluster pairs. (4) REWs measured from the SNR-weighted mean stacked spectra within $\pm$500\kms\ around \Lya\. The values in the parenthesis are the REWs measured from the median stacked spectra.  (5) \& (6)  Line widths and Line centroids obtained from Gaussian fitting in the median stacked spectra.
\end{tablenotes}
\end{table*}

\begin{table*}
\caption{Details of the covering fraction measurements for \Lya, \CIV, and \OVI.}
\label{tab:results_CF} 
\begin{tabular}{cccccccccccc}
    \hline
\multicolumn{1}{c}{Species} & Sample &  \multicolumn{3}{c}{ W$_{\rm th} = $0.1\AA}  & \multicolumn{3}{c}{ W$_{\rm th} = $0.2\AA} & \multicolumn{3}{c}{ W$_{\rm th} = $0.3\AA} \\
                            &         &   \rhocl (Mpc)    & \nrhocl  &   CF           &  \rhocl (Mpc)    & \nrhocl  &   CF           &  \rhocl(Mpc)    & \nrhocl  &   CF  \\ 
\hline
    
\multirow{5}{*}{\Lya}     &Full  &4.89$^{+2.18}_{-1.9}$   &7.1$^{+3.4}_{-2.1}$   &0.21$^{+0.01}_{-0.01}$  &4.88$^{+2.17}_{-1.97}$  &7.1$^{+3.4}_{-2.1}$   &0.19$^{+0.01}_{-0.01}$  
                          &4.88$^{+2.17}_{-1.96}$  &7.1$^{+3.4}_{-2.1}$   &0.16$^{+0.01}_{-0.01}$\\
                         &Bin1  &0.56$^{+0.26}_{-0.55}$  &0.7$^{+0.4}_{-0.4}$   &0.27$^{+0.08}_{-0.09}$  &0.58$^{+0.32}_{-0.47}$  &0.8$^{+0.5}_{-0.5}$  &0.26$^{+0.07}_{-0.08}$  &0.60$^{+0.32}_{-0.43}$   &0.8$^{+0.5}_{-0.5}$  &0.19$^{+0.06}_{-0.07}$\\
                         &Bin2  &2.85$^{+1.01}_{-0.98}$  &4.1$^{+1.6}_{-1.1}$   &0.26$^{+0.03}_{-0.03}$  &2.87$^{+1.0}_{-1.01}$   &4.1$^{+1.5}_{-1.1}$  &0.23$^{+0.02}_{-0.02}$  &2.87$^{+1.0}_{-1.01}$   &4.1$^{+1.5}_{-1.1}$  &0.22$^{+0.02}_{-0.02}$\\
                         &Bin3  &4.99$^{+0.99}_{-1.26}$  &7.1$^{+1.1}_{-0.7}$   &0.21$^{+0.02}_{-0.03}$  &4.96$^{+0.93}_{-1.13}$  &7.1$^{+1.1}_{-0.7}$  &0.19$^{+0.02}_{-0.02}$  &4.96$^{+0.93}_{-1.12}$  &7.1$^{+1.1}_{-0.7}$  &0.16$^{+0.02}_{-0.02}$\\
                         &Bin4  &6.48$^{+1.08}_{-1.66}$  &9.2$^{+0.5}_{-0.5}$   &0.16$^{+0.02}_{-0.02}$  &6.52$^{+1.12}_{-1.6}$   &9.2$^{+0.6}_{-0.5}$  &0.15$^{+0.02}_{-0.02}$  &6.52$^{+1.12}_{-1.57}$  &9.2$^{+0.6}_{-0.5}$  &0.12$^{+0.02}_{-0.02}$\\\\
\hline
                            &        &  \multicolumn{3}{c}{ W$_{\rm th} = $0.05\AA}  & \multicolumn{3}{c}{ W$_{\rm th} = $0.1\AA} & \multicolumn{3}{c}{ W$_{\rm th} = $0.15\AA} \\
                           &         &   \rhocl(Mpc)    & \nrhocl  &   CF           &  \rhocl(Mpc)    & \nrhocl  &   CF           &  \rhocl(Mpc)    & \nrhocl  &   CF  \\

\hline
\multirow{4}{*}{\CIV} &Full     &4.39$^{+1.77}_{-2.11}$  &6.1$^{+2.4}_{-2.8}$   &0.10$^{+0.03}_{-0.04}$   &5.32$^{+2.5}_{-1.9}$          &6.8$^{+2.7}_{-2.6}$   &0.08$^{+0.02}_{-0.02}$ 
                                &5.06$^{+2.23}_{-2.27}$  &6.7$^{+2.6}_{-2.6}$   &0.06$^{+0.01}_{-0.01}$ \\
                                &Bin1  &2.71$^{+1.43}_{-0.63}$  &3.7$^{+1.9}_{-0.8}$   &0.07$^{+0.03}_{-0.06}$  &2.91$^{+1.11}_{-0.99}$  &4.1$^{+1.8}_{-1.1}$  &0.05$^{+0.02}_{-0.03}$ &2.89$^{+1.41}_{-1.07}$  &4.1$^{+1.9}_{-1.1}$  &0.05$^{+0.02}_{-0.02}$ \\
                                &Bin2  &4.34$^{+0.93}_{-2.04}$  &6.1$^{+0.9}_{-0.7}$   &0.18$^{+0.06}_{-0.08}$  &5.32$^{+1.22}_{-1.14}$  &6.8$^{+0.7}_{-0.8}$  &0.12$^{+0.03}_{-0.04}$ &5.09$^{+1.05}_{-1.36}$  &6.7$^{+0.7}_{-1.0}$  &0.08$^{+0.02}_{-0.03}$ \\
                                &Bin3  &6.28$^{+0.87}_{-1.38}$  &8.9$^{+1.5}_{-0.8}$   &0.07$^{+0.03}_{-0.06}$  &6.91$^{+1.01}_{-1.62}$  &9.3$^{+0.9}_{-0.5}$  &0.06$^{+0.02}_{-0.03}$ &6.85$^{+0.96}_{-1.89}$  &9.3$^{+0.8}_{-0.5}$   &0.05$^{+0.02}_{-0.02}$ \\\\

\multirow{4}{*}{\OVI} &Full    &4.79$^{+2.12}_{-1.70}$   &6.9$^{+3.2}_{-2.2}$   &0.10$^{+0.01}_{-0.02}$   &4.88$^{+2.2}_{-1.85}$        &7.1$^{+3.3}_{-2.0}$   &0.05$^{+0.01}_{-0.01}$ 
                               &4.83$^{+2.19}_{-1.94}$  &7.0$^{+3.3}_{-2.1}$   &0.04$^{+0.01}_{-0.01}$  \\
                               &Bin1 &2.70$^{+1.14}_{-0.97}$   &3.8$^{+1.9}_{-1.4}$   &0.12$^{+0.02}_{-0.03}$  &2.71$^{+1.32}_{-0.89}$  &3.4$^{+2.1}_{-1.2}$  &0.06$^{+0.01}_{-0.01}$ &2.69$^{+1.23}_{-0.90}$   &3.9$^{+1.9}_{-1.2}$  &0.04$^{+0.01}_{-0.01}$  \\
                               &Bin2 &4.95$^{+1.02}_{-1.12}$  &6.9$^{+0.7}_{-0.8}$   &0.10$^{+0.02}_{-0.03}$   &4.95$^{+1.01}_{-1.05}$  &7.1$^{+1.0}_{-0.7}$  &0.05$^{+0.01}_{-0.01}$ &4.88$^{+0.95}_{-1.12}$  &7.0$^{+1.0}_{-0.8}$  &0.05$^{+0.01}_{-0.01}$  \\
                               &Bin3 &6.19$^{+0.96}_{-1.56}$  &9.1$^{+0.6}_{-0.6}$    &0.07$^{+0.02}_{-0.02}$  &6.43$^{+1.02}_{-1.66}$  &9.1$^{+0.6}_{-0.6}$  &0.04$^{+0.01}_{-0.01}$ &6.43$^{+0.98}_{-1.7}$   &9.1$^{+0.6}_{-0.6}$  &0.03$^{+0.01}_{-0.01}$  \\

    \hline
\end{tabular}
\label{tab:multicol}
\end{table*}

\begin{figure*}
\begin{center}
     \includegraphics[width=1.1\textwidth]{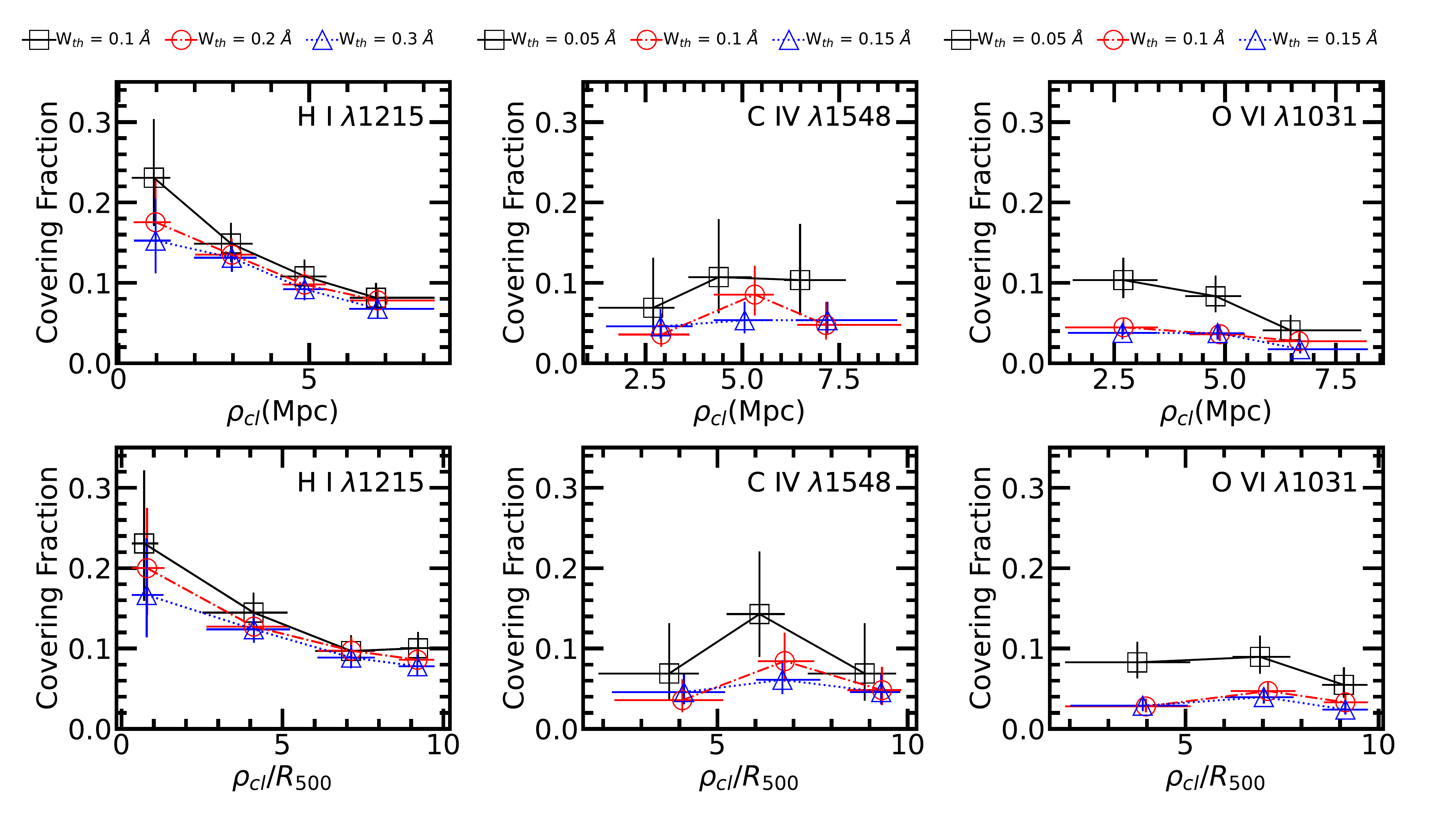}
     \caption{Same as Fig.~\ref{fig:covering_fraction1} but for the Flag-1 absorption systems.}
     \label{fig:covering_fraction2}
\end{center}
\end{figure*}

\section{Comparative Analysis of Stacking Methods}
\label{appendix:snr_effect}

To investigate the effects of varying SNR of the individual spectra in the REW measurements from the composite spectra, we generate mock  stacked spectra for 5 different SNR per pixel bins ranging from 0 to 25 (i.e., 0--5, 5--10, 10--15, 15--20, and 20--25). Note that the COS spectra used in our study have SNRs typically in this range. For a given SNR bin, we first produce 1000 synthetic normalized COS spectra. We then add Gaussian noise to each spectrum corresponding to the SNR of the respective bin by drawing a uniform random number within the SNR range.

\begin{figure*}
\begin{center}
     \includegraphics[width=1.0\textwidth]{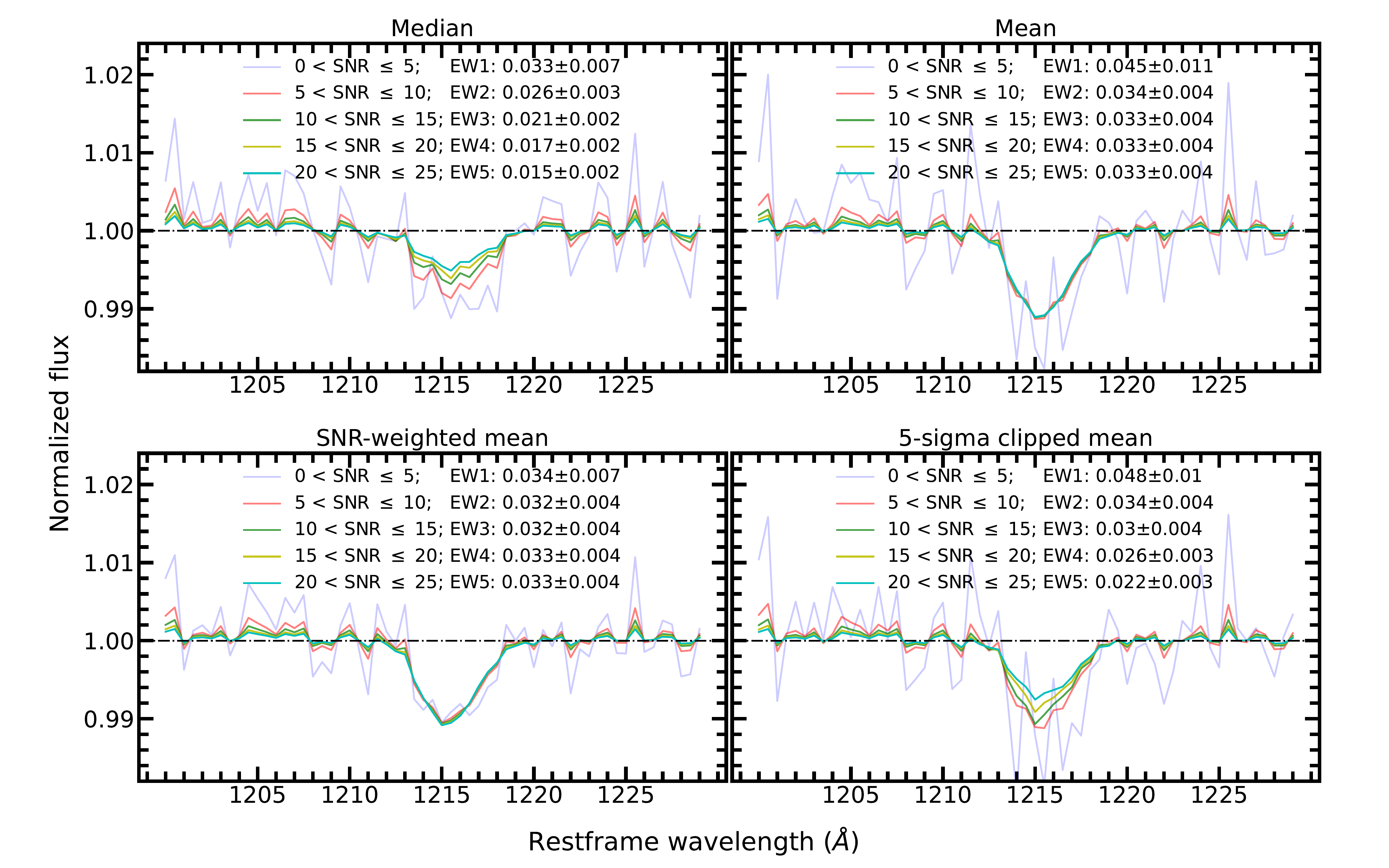}
     \caption{ Comparison of mock-stacked \Lya\ absorption profiles for five SNR bins, as indicated by the legends. {\tt Top-left:} Median-stacked spectra. Each spectrum for each SNR bin is obtained using the median stacking of 1000 spectra with randomly injected \Lya\ absorbers with REW $>0.1$~\AA, and a covering fraction of 0.21, and a velocity dispersion of 436~\kms\ around the \Lya\ wavelength. The measured REWs for different SNR bins are indicated in the legends. The errors in REWs are estimated using 200 bootstrap realizations of the 1000 input spectra. {\tt Top-right:} The same as {\tt top-left} but for the mean statistic. {\tt Bottom-left:} The same as {\tt top-left} but for the SNR-weighted mean statistic. {\tt Bottom-right:} The same as {\tt top-left} but for the 5-sigma clipped mean statistic. Evidently, the SNR-weighted mean statistic is insensitive to the SNR of the individual input spectra.}  
     \label{fig:snr_effect_mock}
\end{center}
\end{figure*}

Next, based on the observed \Lya\ CF of 0.21 (see Section~\ref{subsec:covering_fraction}) for $W_{\rm th}$ > 0.1\AA\ absorption systems, we randomly inject 210 \Lya\ absorbers with $W_{\rm th}$ > 0.1\AA\ in these 1000 spectra. We distribute these absorbers around the rest wavelength of \Lya\ with the observed velocity dispersion of 436~\kms (see Fig.\ref{fig:stackprofiles}). We use Gaussian profiles to model these absorbers using the REW and b values of the \Lya\ absorbers that are randomly drawn from \citet{Danforth2016}. We then stack these 1000 spectra using four statistical methods: median, mean, SNR-weighted mean, and $5\sigma$-clipped mean. The resulting stacked profiles are compared in Fig.~\ref{fig:snr_effect_mock}.

The REWs are measured within $\pm$500 km/s around \Lya, and the corresponding $1\sigma$ uncertainties are derived from 200 bootstrap realizations of the 1000 input spectra for each SNR bin. From Fig.~\ref{fig:snr_effect_mock}, it can be seen that the derived REWs are sensitive to the SNR of the input spectra for the median-stacked profiles, with the REW gradually increasing with decreasing SNR. A similar feature is also seen for the $5\sigma$-clipped mean stacks. For the mean stack, the REW value is converged for all but the lowest SNR bin. The SNR-weighted mean profiles, however, have similar REWs across all the SNR bins, suggesting that this method is insensitive to the SNR of the input spectra. We therefore adopted the SNR-weighted mean for our analysis to minimize the effects of varying SNR of individual quasar spectra.

\section{Mock Analysis of Weak \Lya\ absorption}
\label{appendix:mock_weak_abs}

\begin{figure*}
\begin{center}
     \includegraphics[width=1.1\textwidth]{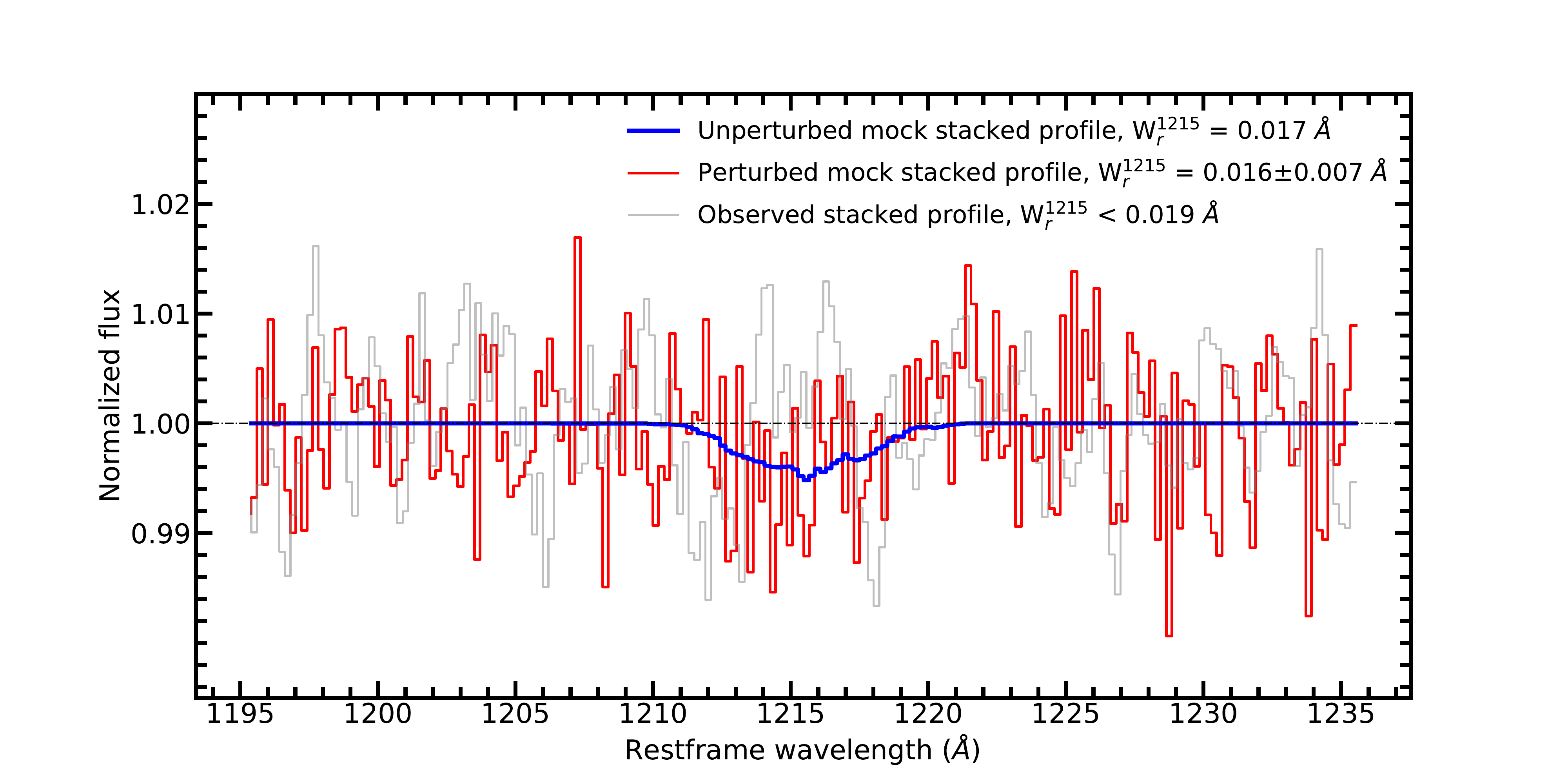}
     \caption{ Mean composite spectrum  (\emph{blue}) generated by stacking 942 mock spectra without any noise. Detailed information regarding the generation of the stack is provided in Section~\ref{appendix:mock_weak_abs}. The same mean stacked spectrum perturbed with a Gaussian noise corresponding to an SNR of 150 per pixel is shown in red. This SNR value corresponds to the SNR in the line-free region of the SNR-weighted mean stacked profile of the 942 quasar-cluster pairs for our sample (in gray). The REWs estimated within $\pm$500\kms\ around the \Lya\ for both unperturbed and perturbed mock-stacked spectra, as well as the $3\sigma$  upper limit from the observation, are displayed at the top right corner. The $1\sigma$ scatter in the REW is determined by repeating the analysis 200 times.}

     \label{fig:mock_simulation}
\end{center}
\end{figure*}

To understand the non-detection of \Lya\ absorption signal in the composite spectrum in Section~\ref{subsec:infall_substructure} when the strong absorbers ($N_{\rm HI} > 10^{13.3}\rm~cm^{-2}$) are excluded, we perform a mock stacking analysis using synthetic weak \Lya\ absorbers (10$^{12.0}$ < $N_{\rm HI}$ (\sqcm) < 10$^{13.3}$). To mimic the observation, we produced 942 synthetic COS spectra centered on 1215.67~\AA. Next, we inject \Lya\ absorbers with 10$^{12.0}$ < $N_{\rm HI}$ (\sqcm) < 10$^{13.3}$ in these spectra randomly, assuming the column density distribution function of \citet{Danforth2016}. \footnote{For simplicity, we are assuming that the number density of \Lya\ absorbers is the same around clusters and random regions (IGM). The clustering of \Lya\ absorbers with galaxy clusters would increase the number of actual absorbers.} For each absorber, we randomly choose a $b$-value from the $b$-distribution for its column density bin from \citet{Danforth2016}. The REW of an absorber is estimated from its column density assuming the linear part of the curve of growth (COG), which is a good approximation for $N(\HI) < 10^{13}$~\sqcm. With these $b$ and REW values, we insert Gaussian absorption profiles with centorids drawn from a Gaussian distribution with a $\sigma$ of 436~\kms\ (i.e., the $\sigma$ of the stacked profile; Fig.~\ref{fig:stackprofiles}) with respect to the \Lya\ line centre. If a sightline includes multiple absorbers of varying column densities, the final spectrum is generated by combining the contributions from each absorber. After inserting the absorbers, we convolve each sightline with a Gaussian kernel of FWHM of 17.7~\kms, mimicking the spectral resolution of COS. We also rebin each spectrum with 3-pixels using {\it SpectRes} \citep{2017arXiv170505165C}, to ensure consistency with the observation. We then perform a mean stack of these 942 spectra. This entire analysis is repeated 200 times, and the mean of all the mean-stacked spectra serves as our final unperturbed stacked spectrum and is shown in blue in Fig.~\ref{fig:mock_simulation}. Next, we introduce Gaussian noise to perturb the final stacked spectrum so that the SNR per pixel of the resultant spectrum is consistent with the observed value (i.e., 150 per pixel). 
The perturbed stacked profile is shown in red in Fig.~\ref{fig:mock_simulation}. We estimate REWs of $0.017$~\AA\ and $0.016\pm0.007$~\AA\ (not formally detected with $>3\sigma$) for the unperturbed and perturbed mean stacked profiles, respectively. The $3\sigma$ upper limit on REW obtained from the observed stacked spectrum is 0.019~\AA\ (see Section~\ref{subsec:infall_substructure}). Clearly the expected \Lya\ absorption signal is lower than this $3\sigma$ limit. It becomes visually evident from Fig.~\ref{fig:mock_simulation} when the unperturbed stack is overlaid with noise comparable to the observation. We found that a spectral SNR of 200 per pixels is required to detect the expected  signal with $>3\sigma$ confidence.

\label{lastpage}
\end{document}